\documentclass[3p]{elsarticle}

\usepackage[utf8]{inputenc}
\usepackage[USenglish]{babel} 
\usepackage[T1]{fontenc}
\usepackage{lmodern} 
\usepackage{color}
\usepackage{mathtools}
\usepackage{graphicx} 

\usepackage{setspace}

\usepackage{amsmath}
\usepackage{amsthm}
\usepackage{amsfonts}

\newcommand{\bx}{{\bf x}}
\newcommand{\q}[2]{\hat{q}_{#1}\!  \left(#2 \right) }
\newcommand{\shat}{\hat{s}}

\newcommand{\phat}{\hat{p}}

\newcommand{\Ha}[2]{\mathrm{H}_{#1}\!  \left(#2 \right) }

\newcommand{\I}{\mathrm{i}}
\newcommand{\E}{\mathrm{e}}
\newcommand{\Om}{{\omega_0}}
\newcommand{\Fm}{{f_0}}

\newcommand{\dint}{\int\limits_{-\infty}^{\infty} \! \int\limits_{-\infty}^{\infty}}
\newcommand{\sint}{\int\limits_{-\infty}^{\infty} }

\newcommand{\Ompm}[1]{{\omega_{#1}'}}

\newcommand{\Fmpm}[1]{{f_{#1}'}}

\journal{J. Sound Vib.}
\begin{document}
\begin{frontmatter}
\author{Christian H. Kasess\fnref{ARI}}
\ead{christian.kasess@oeaw.ac.at}
\author{Wolfgang Kreuzer\fnref{ARI}}
\author{Prateek Soni\fnref{ARI}}
\author{Holger Waubke\fnref{ARI}}

 \affiliation[ARI]{organization={Acoustics Research Institute, Austrian Academy of Sciences},
             addressline={Dominikanerbastei 16},
             city={Vienna},
             postcode={A-1010},
             country={Austria}}

\cortext[cor1]{Corresponding author: Christian H. Kasess}

\title{Localizing uniformly moving single-frequency sources using an inverse 2.5D approach}

\begin{abstract}
The localization of linearly moving sound sources using microphone arrays is particularly challenging as the transient nature of the signal leads to relatively short observation periods. Commonly, a moving focus approach is used and most methods operate at least partially in the time domain. In contrast, this manuscript presents an inverse source localization algorithm for uniformly moving single-frequency sources that acts entirely in the frequency domain. For this, a 2.5D approach is utilized and a transfer function between sources and a microphone grid is derived. By solving a least squares problem using the measured data at the microphone grid, the unknown source distribution in the moving frame can be determined. First, the measured time signals need to be transformed from the time into the frequency domain using a windowed discrete Fourier transform (DFT), which leads to an effect called spectral leakage that depends on the length of the time interval and the analysis window used. 

To include the spectral leakage effect in the numerical model, the calculation of the transfer matrix is modified using the Fourier transform of the analysis window used in the DFT applied to the measurements. Currently, this approach is limited to single-frequency sources as this restriction allows for a simplification of the calculation and reduces the computational effort. The least squares problem is solved using a Tikhonov regularization employing an L-curve approach to determine a suitable regularization parameter. As moving sources are considered, utilizing the Doppler effect enhances the stability of the system by combining the transfer functions for multiple frequencies in the measured signals. The performance of the approach is validated using simulated data of a moving point source with or without a reflecting ground. Numerical experiments are performed to show the effect of the choice of frequencies in the receiver spectrum, the effect of the DFT, the frequency of the source, the distance between source and receiver, and the robustness with respect to noise. 
\end{abstract}

\end{frontmatter}

\section{Introduction}
The localization of sound sources is an important topic in order to efficiently reduce noise burden in the environment. Microphone arrays are one of the major tools used for sound source localization. 
By far the most popular method is beamforming and all its flavors (see, e.g. \cite{Chiariotti2019,Merino2019} for extensive reviews on the topic). 

Briefly, beamforming methods virtually focus the microphone array on a potential source position, suppressing signals coming from other locations. In general, these methods are based on the assumption of free-field wave propagation.
In stationary conventional beamforming, where non-moving source are assumed, the cross-spectral matrix of the microphone signals is calculated as well as the so-called steering vector. The steering vector determines the focus of the beamforming algorithm based on the known propagation of the signal from the respective source position to the microphone positions. Using all points of a predefined source grid as focus points, a beamforming map is generated.  
The so produced raw beamforming map usually is subject to a point-spread-function, i.e. the beamforming map of a point source is blurred and subject to sidelobes depending on different factors such as the frequency, array geometry, as well as distance between the array and source grid. To produce a better resolved source map, deconvolution approaches such as DAMAS (deconvolution approach for the mapping of acoustic sources, \cite{Brooks2006,Dougherty2005}) or CLEAN-SC (CLEAN based on source coherence, \cite{Sijtsma2007}) are applied to the beamforming results. For a comparison of different deconvolution methods see, e.g.~\cite{Herold2017} and for more details on beamforming in general please refer to recent reviews on the topic, e.g.  \cite{Chiariotti2019,Merino2019}.

For moving sources, the typical approach is to use a moving focus, i.e. a steering vector which is altered over time. The time-dependent alignment of the array is used to compensate for the position dependent Doppler effect. This technique is referred to as de-Dopplerization \cite{Howell1986} and it is an essential part of beamforming approaches for moving sources (see e.g. \cite{Cousson2019,Fleury2011,Guerin2008,Zhang2019,Zhang2023}). In \cite{Fleury2011}, a frequency domain approach was developed where short time windows are used to limit the source displacement allowing for an approximate Doppler compensation. Various deconvolution approaches were then combined with this spectral beamforming technique. Methods which are at least partially located in the time domain are used more commonly for moving sources (cf. \cite{Cousson2019,Guerin2008,Schumacher2022,Zhang2019,Zhang2023}). These methods are based on a time domain formulation for the pressure caused by a moving point source which is used to de-Dopplerize the time signal directly. As the de-Dopplerization is applied at every time step, 
no assumptions about small displacements are necessary. However, in the time domain the propagation path must typically be calculated for a very high number of time points. While the method proposed in \cite{Cousson2019} works fully in the time domain, many approaches apply frequency domain methods after some initial time domain processing steps involving de-Dopplerizing the original time signal (e.g. \cite{Guerin2008,Schumacher2022,Zhang2019,Zhang2023}). 
Importantly, the present work is solely concerned with linearly moving sources. Other types of motion may require other approaches, see e.g. \cite{Chu2021,Gombots2021} for rotating fans.

Irrespective of whether the time or the frequency domain is used or which type of deconvolution approach is applied, the aforementioned methods and their many variants are based on conventional beamforming, i.e. focusing the array using a steering vector. 
A different idea is used in so-called inverse schemes. In conventional beamforming, the strength of each source is determined independently of the others. Only the deconvolution step, which also solves an inverse problem, considers the dependencies within this raw beamforming map. In contrast, inverse methods exist, where a joint optimization problem is solved directly on the measurements. The remainder of this manuscript is concerned with this latter class of methods. 
Although these inverse methods can of course be applied to the free-field wave propagation (e.g. generalized inverse beamforming \cite{Suzuki2011}, spectral estimation methods \cite{Blacodon2004,Yardibi2010,Herold2017} and extensions thereof \cite{Oertwig2022}, compressive beamforming \cite{Meng2019}), the inverse approach allows for a much more general definition of sources and the surrounding environment. Using numerical approaches like the boundary element method (BEM) or the finite element method (FEM), scattering structures and other types of sources such as velocity elements can be included in the forward model (e.g. \cite{Gombots2021,Schuhmacher2003}) which is not or only to a limited degree possible in conventional beamforming (including deconvolution). This is, e.g. important when significant ground reflections affect the measurements, or in confined spaces when reflecting objects are in the vicinity (e.g. \cite{Gombots2021}). 
Note that for the case of point sources over a fully reflecting ground, a conventional beamforming approach was also developed \cite{Christensen2004} which was later extended to include a deconvolution step \cite{Zhigang2016}.

Although computationally demanding,  the forward problem is often comparatively straight forward for stationary sources. However, the inversion of the forward map poses some difficulties. As there are, in general, more potential source positions than receiver points (i.e. microphones), the transfer map/matrix from source to array is underdetermined. Many different approaches exist to address this problem, and the way the inverse problem is regularized allows the control of properties of the solution such as smoothness or sparseness. 
For example, in \cite{Schuhmacher2003}, a standard Tikhonov regularization was used to stabilize the inversion, which led to typically smooth source maps. To find a suitable regularization parameter, the L-curve approach was applied, where a range of regularization parameters is used to solve the problem and from these results the degree of regularization is determined that ``best'' represents a compromise between model error and regularization (for details see e.g. \citep{Hansen1993}). 
If it can be assumed that only few localized source points have non-zero strengths, sparse methods, that, e.g.  minimize the $L_1$ norm of the solution instead of the $L_2$ norm, can be used (cf. generalized inverse beamforming, \cite{Suzuki2011}). Recently, an inverse scheme based on FEM for stationary sources was introduced \cite{Gombots2021} where a norm was used on the source map coefficients with a non-integer exponent lying between 1 and 2. When setting the value close to 1, some sparsity can be enforced. Another example of an inverse method is given in \cite{Meng2019}, where a compressive beamforming approach using an $L_1$ regularization is introduced for linearly moving sources. The method operates entirely in the time domain with the forward problem being set up for time points in a pre-defined period and then the full system is solved.  These are only but a few methods to find regularized solutions to linear systems of equations.

In the present work, an inverse approach for sound source localization of single-frequency sources moving at a constant speed is presented that operates fully in the frequency domain. 
The effect of the motion of the source is incorporated in closed form in the forward transfer function using a 2.5D approach based on the Helmholtz equation. This allows for the consideration of, in theory, arbitrarily long time windows without the necessity of dividing the observations into shorter segments.
In the 2.5D approach a constant cross-section in the $y$-$z$ plane is assumed, but in contrast to pure 2D methods the sound field may vary along $x$, allowing for the modeling of point sources, incoherent line sources, or sources of finite length. This is achieved by employing a spatial Fourier transform along $x$ to transform the Helmholtz equation into the wavenumber domain. The 3D sound field is calculated by solving a number of 2D problems with different wavenumbers and an inverse Fourier transform is used to get back to the spatial domain in $x$ (cf.  \cite{Duhamel1996,Fakhraei2022,Hornikx2007,Kamrath2018,Kasess2016,Li2020,Pizarro-Ruiz2019,Wei2021}).  
The 2.5D BEM also allows for a straightforward definition of impedance boundary conditions \cite{Duhamel1998}. Thus, the Helmholtz BEM including its 2.5D variant is a very versatile and established tool in noise research.

A major advantage of the 2.5D approach is that sources moving along $x$ at a constant speed $v_s$ can be treated in a straightforward manner  (see e.g. \cite{Duhamel1996}). However, the main problem of this approach is that the measured signals at the microphone array and the numerically calculated transfer matrix, which consists of the transfer functions from source to the receiver points, are defined in different domains, i.e. the time and the frequency domain, respectively. Thus, the first step is to transform them into a common domain, which is in our approach given by the frequency domain.

Unfortunately, while the transfer functions and hence the transfer matrix used for the inversion are described using \emph{continuous} Fourier integral transforms, the measured data can only be transformed using a \emph{discrete} Fourier transform (DFT).  This transformation results in undesirable effects, most notably spectral leakage from adjacent frequency bands which is dependent on the length of the observation period and the possible use of an analysis window before the DFT. While the choice of the latter allows some control over the leakage, increasing the observation period is very limited in the case of linearly moving sources, because pass-by events are, in general, very brief. As a consequence, even though the transfer function and the measurements after applying the DFT are both in the frequency domain, there is a substantial difference between the two entities.
The novelty in the approach presented is to include the effects of the DFT in the calculation of the final transfer matrix, thus bringing it to the same discrete frequency domain as the measurements. 
The inversion of this transfer matrix is done using a Tikhonov regularization and the regularization factor is determined using the L-curve approach \cite{Hansen2007}. To stabilize the L-curve, a small amount of uncorrelated noise was used as suggested in \citep{Johnston2000}.
This basic approach is chosen because the focus of the present work is to investigate the general mapping properties of the matrix with respect to the achievable resolution for point sources. Finding potentially better suited inversion routines based , e.g. on compressive sensing techniques, are beyond the focus of this manuscript.

The Helmholtz equation is derived using a harmonic ansatz. As a consequence, inverse methods based on the Helmholtz equation assume the source signals to be a sum of complex exponentials (e.g. \cite{Gombots2021,Schuhmacher2003}). While for a stationary source these complex exponentials can be treated independently, this independence is not given in case of a moving source, which results in a high computational effort.  To allow for a simplification of the model and for a reduction in the computational effort, the algorithm is currently limited to single-frequency sources of constant amplitude. Even this restricted setting already illustrates the general principle of our proposed algorithm. 
From a practical point-of-view the treatment of multiple frequency components simultaneously is, of course, of great interest. By using a joint inversion across all frequencies of interest it is possible to extend the algorithm to include multiple frequency components. This, however, leads to a much larger transfer matrix  and an increase in computational complexity concerning both the calculation of the transfer matrix and its inversion. Finding ways to reduce the effort is ongoing work and will not be covered here.

The newly developed algorithm was thoroughly tested using simulated data comprising moving single-frequency point sources in the free-field with and without a reflecting half-plane to model reflections from the ground. The test data was generated using a time domain approach (cf. \cite{DeHoop2005}). 
A state-of-the-art array geometry based on a spiral arrangement was used to define the microphone positions. The effect of different parameters such as source frequency, analysis window length, and choice of observed frequencies was investigated. 

This paper is structured as follows. First, a detailed derivation of the algorithm for a general 2.5D boundary element setting is provided in Sec.~\ref{sec:meth}.  After an overview of the simulated test cases, the results on the method's performance under different settings are provided in Sec.~\ref{Sec:Results}.  Note that the test cases used in this work comprise only moving point sources, the calculation of which does not involve actual boundary element models. Although the method itself is valid for a general BEM setting involving scattering objects, these examples were chosen to simplify and to speed up the calculation of the acoustic field at the microphone. Nevertheless, in spite of their simplicity, the examples illustrate important properties of the proposed algorithm. A detailed discussion of the results as well as concluding remarks are provided in Sec.~\ref{sec:disc} and Sec.~\ref{sec:concl}, respectively.

\section{Methods}
\label{sec:meth}

\subsection{Transfer function in 2.5D}
The 2.5D approach was developed to efficiently calculate scattering from very long structures in 3D for stationary as well as for uniformly moving sources~\cite{Duhamel1996}. In this section, a concise introduction to this method is given to keep the manuscript self contained.

If the scattering object is infinitely long along one direction and has a constant cross-section (in practice this is, e.g. approximately true for a long noise barrier), the 2.5D approach can be used to reduce a full 3D acoustic scattering problem into a series of 2D problems with different wavenumbers. For the remainder of this manuscript the infinite dimension, which is also the direction of motion, is defined along the $x$-axis, while the 2D cross-section is in the $y$-$z$-plane.  

Briefly, the general 2.5D approach is based on a Fourier transform $x \xrightarrow{\mathcal{F}} k_x$ (cf.~\cite{Duhamel1996,Fakhraei2022,Hornikx2007,Kasess2016,Li2020,Pizarro-Ruiz2019,Wei2021}), where $k_x$ is the wavenumber with respect to the $x$-axis. 
With this transformation, the Green's function for the Helmholtz equation in 3D can be represented by  
\begin{align}\label{Equ:Greens25D}
\frac{\E^{\I k r_{3}}}{4 \pi r_{3}} &= \frac{1}{2 \pi} \sint \frac{\I}{4}  \Ha{0}{r_2 \sqrt{\omega^2/c^2-k^2_x}}  \E^{\I k_x (x_r-x_s)} dk_x ,
\end{align}
where $\bx_s=\left(x_s,y_s,z_s\right)^\top$ and $\bx_r=\left(x_r,y_r,z_r\right)^\top$ are a given source and receiver point in 3D, respectively (cf. Eq.~(6.616) in \cite{Gradshteyn6} or Eq.~(5) in \cite{Duhamel1996}). $r_2 = \sqrt{ (y_r - y_s)^2 + (z_r - z_s)^2}$ is defined as the Euclidean distance in the $y$-$z$-plane between these points and $r_{3} = ||\bx_r - \bx_s||_2$ gives their distance in 3D. $\Ha{0}{\cdot}$ denotes the Hankel-function of the first kind of order 0 (the Green's function for the 2D Helmholtz equation), $ k = \omega c^{-1}$, where  $\omega$ is the angular frequency of interest for the 3D problem and $c$ denotes the speed of sound. In short, the 3D Green's function with wavenumber $k = \omega c^{-1}$ can be represented by an integral with respect to $k_x$ (inverse Fourier transform) over 2D Green's functions with 2D wavenumbers $k_2=\sqrt{\omega^2/c^2 - k_x^2}$. $k_x$ denotes the (spatial) wavenumber along the $x$-dimension. Note that in this manuscript the sign convention used in \cite{Duhamel1996}, i.e. the harmonic ansatz $\E^{- \I \omega t}$, is followed.

More generally, this approach can be combined with the boundary element method (BEM) to include (infinitely long) scatterers with constant cross-sections in the $y$-$z$ plane (for a derivation see e.g.~\cite{Duhamel1996}). 
In this case, the sound pressure at a point $\bx_r$ in 3D 
can be calculated as an inverse Fourier transform: 
\begin{align}\label{Equ:25D}
\phat\left(x_s,y_s,z_s,x_r,y_r,z_r,\omega\right) &= \frac{1}{2 \pi} \sint \q{}{y_s,z_s,y_r,z_r,\sqrt{\omega^2/c^2-k^2_x}}  \E^{\I k_x (x_r-x_s)} dk_x ,
\end{align}
with $\q{}{\cdot}$ being the 2D Helmholtz BEM solution at the receiver point in the $y$-$z$-plane for a source located at $\left(y_s,z_s\right)^\top$ and a 2D wavenumber $k_2=\sqrt{\omega^2/c^2-k^2_x}$. With a slight abuse of notation $\hat{f}$ denotes the Fourier transform of a function $f$ with respect to either time or space or both. It is indicated by the argument which Fourier transformation was used. $x_r-x_s$ simply denotes the offset of the receiver with respect to the source position along the $x$-dimension. Importantly, the term source does not imply a specific source type and could be an external point source but also velocity or pressure boundary conditions, as long as their Fourier transform with respect to $x$ exists. For details on the definition of velocity boundary conditions in the 2.5D BEM see \cite{Li2020}. In addition to different source types, the 2.5D BEM can also include admittance boundary conditions and absorbing grounds \cite{Duhamel1998}.
Eq.~(\ref{Equ:25D}) can also be defined for multiple concurrent sources. However, if the sources have different $x$-offsets, their respective contributions to $\q{}{\cdot}$ either have to be weighted with the appropriate phase factor resulting from the spatial shift or calculated separately and then added. 

For the simplest case of a single point source at $\bx_s = (x_s,y_s,z_s)^\top$ and no scattering objects, the 2D solution is simply $\q{}{y_s,z_s,y_r,z_r,k_2} = \frac{\I}{4} \Ha{0}{r_2 k_2}$ as in Eq.~(\ref{Equ:Greens25D}). 

To numerically calculate the inverse Fourier transform, the integral in Eq.~(\ref{Equ:25D}) is replaced by a quadrature formula, e.g. Gauss-type quadrature (\cite{Duhamel1996,Wei2021}) or a Filon-Clenshaw-Curtis quadrature \cite{Kasess2016}. For every quadrature node one 2D calculation needs to be performed. 

\subsubsection{Source-receiver relation for a uniformly moving source}
Eq.~(\ref{Equ:25D}) is only valid for stationary sources. The time-dependent pressure caused by a uniformly moving point source, which is the case investigated here, was derived previously (Eqs.~(43)-(50) in \cite{Duhamel1996}). Briefly, a single point source emitting a source signal $s(t)$ in the time domain is assumed which moves at a constant speed of $v_s$ along $x$: $x_s(t) = x_{0} + v_s t$. 
The pressure at the receiver position $\left(x_r,y_r,z_r\right)^\top$ at time $t$ is given as:

\begin{align}\label{Equ:Uni}
p(x_{0},y_s,z_s,x_r,y_r,z_r,t) &= \frac{1}{(2 \pi)^2} \dint  \shat(\omega - v_s k_x) \q{}{y_s,z_s,y_r,z_r,\sqrt{\omega^2/c^2-k^2_x}}  \E^{\I k_x (x_r - x_{0})} \E^{-i \omega t} dk_x  d\omega.
\end{align}
$\shat(\omega - v_s k_x)$ is the Fourier transform of $s(t)$ evaluated at $\omega - v_s k_x$. In contrast to Eq.~(\ref{Equ:25D}), Eq.~(\ref{Equ:Uni}) defines the pressure at the receiver point $\bx_r$ in the time domain. At this point, it is important to note that there are two commonly used physically different definitions for a moving point source. The definition used here defines a moving point source in the wave equation for sound pressure. Alternatively, a point source can be defined in the velocity potential wave equation which is equivalent to a small pulsating sphere (e.g. \cite{Morse1987}). This results in a temporal derivative on the right hand side of the pressure wave equation which can also be handled in the 2.5D approach.
The definition in the time domain provides the basis for incorporating the effects of the \textit{discrete} Fourier transform (DFT) into the transfer matrix.   

A general source localization scenario is defined by a (possibly moving) source region containing $L$ potential (point) sources at positions $\bx_{s,\ell}, \ell = 1,\dots, L$ and a microphone array with $N$ microphones at positions $\bx_{r,n}, n = 1,\dots, N$, and typically $N<L$. In the case of a moving point source along the $x$-axis, the first component $x_{s,\ell}$ of $\bx_{s,\ell}$ denotes the $x$-coordinate for the source at time $t=0$\,s. To simplify the notation in Eq.~(\ref{Equ:Uni}), the function arguments relating to a receiver at $\bx_{r,n}$ and to a source at $\bx_{s,\ell}$  will be replaced by function indices $n$ and $\ell$, respectively. With this notation the sound pressure in the time domain is given by 
\begin{align}
\label{Equ:UniInd} 
 p_{n\ell}(t) &= \frac{1}{(2 \pi)^2} \dint  \shat_\ell(\omega - v_s k_x) \q{\ell n}{\sqrt{\omega^2/c^2-k^2_x}}  \E^{\I k_x (x_{r,n} - x_{s,\ell})} \E^{-\I \omega t} dk_x  d\omega,  
\end{align}
the continuous frequency domain formulation reads as
\begin{align}
\label{Equ:UniIndP} 
\phat_{n\ell}(\omega) = \frac{1}{2\pi} \sint \shat_\ell(\omega - v_s k_x) \q{\ell n}{\sqrt{\omega^2/c^2-k^2_x}}  \E^{\I k_x (x_{r,n} - x_{s,\ell})} dk_x.
\end{align}
This equation is valid for any arrangement of sources and receivers.

In an ideal continuous setting, Eq.~(\ref{Equ:UniIndP}) would act as the starting point of the inverse method, as it provides a transfer function between any source point $\bx_{s,\ell}$ and any receiver point $\bx_{r,n}$ that can be used to compare the calculated field at the receiver with a microphone array measurement at the same position. From this, the unknown source spectra $\hat{s_\ell}$ could, in theory, be derived.
In practice, however, Eq.~(\ref{Equ:UniIndP}) cannot be directly applied, and the discrete nature of the measurements and restrictions of the source spectra need to be considered.

\subsection{Discrete Domain Formulation}
The measurements using a microphone array are discrete in space as well as in time. Due to the limited number of microphones and the practical sizes of microphone arrays, a Fourier transformation of the measured data from the spatial $x$-domain into  $k_x$ is not feasible and will thus not be considered. 

The discrete Fourier transform (DFT) required to transform the measured sampled data from the time to the frequency domain is also not ideal since, in the case of linear motion, the observation time is strongly limited. Depending on the noise level and its duration, which is directly related to the speed of the pass-by of the moving source, the signal of interest will eventually disappear in the noisy background at some point. Extending the time any further will increase the noise level of the measurement without gaining any additional information. Limiting the observation time leads to two effects for the DFT: First, the frequency resolution decreases for shorter observation periods, and, secondly, spectral leakage effects increase. Spectral leakage means that the observed spectral level at a given frequency also contains energy from adjacent frequencies. 
While zero padding (sinc-interpolation in $\omega$) can artificially improve the spectral resolution, the leakage effects are not affected and a direct comparison to the continuously transformed coefficients $\phat_{n\ell}$ from Eq.~(\ref{Equ:UniIndP}) may turn out to be problematic. 
This leads to the main idea of transforming Eq.~(\ref{Equ:UniInd}) to include the same effects of the DFT in time used for the measurements. 
In the next section, the approach of how 
the DFT is applied efficiently to Eq.~(\ref{Equ:UniInd}) without actually calculating $p_{n\ell}(t)$ will be described.

\subsubsection{Discrete Fourier transformation of the transfer function}
For deriving the pressure $p(t)$  in Eqs.~(\ref{Equ:UniInd}) and (\ref{Equ:UniIndP}) the (continuous) Fourier transform is used:
\begin{align}
  \phat(\omega) & = \mathcal{F}(p)(\omega) = \int\limits_{-\infty}^\infty p(t) \E^{\I \omega t}dt \label{Equ:time1}, \\ 
  p(t) &= \mathcal{F}^{-1}(\phat)(t) =  \frac{ 1}{ 2 \pi}  \int\limits_{-\infty}^{\infty}  \phat \left(\omega\right) \E^{-\I \omega t} d\omega. \label{Equ:time2}
\end{align}
To avoid a cluttered notation, the indices indicating source and receiver have been dropped. 

Sampling the continuous pressure signal $p(t)$ with a sampling time of $T_s=f_s^{-1}$ and applying an analysis window function $g$ with finite length $\text{supp}(g) \subseteq [0,T_g)$ with $T_g=N_g T_s$,  the \emph{windowed} DFT for a sampled signal of $N_g$ samples for the $m$-th frequency bin $\omega'_m$ can be written as:
\begin{align}
\nonumber DFT_g(p)[\omega'_m] := \sum\limits_{n=0}^{N_g-1}{ p[t_n]} g[t_n] \E^{\I \omega_m' t_n} &= \frac{ 1}{ 2 \pi}  \int\limits_{-\infty}^{\infty}  \phat\left(\omega\right) \sum\limits_{n=0}^{N_g-1}{\!\! \E^{-i \omega t_n} g[t_n] \E^{\I \omega_m' t_n} }\ d\omega\\
\label{Equ:STFT1}&= \frac{ 1}{ 2 \pi}  \int\limits_{-\infty}^{\infty}  \phat\left(\omega\right) \sum\limits_{n=0}^{N_g-1}{ \!\! \E^{\I t_n (\omega_m' - \omega)} g[t_n] }\ d\omega,
\end{align}
where Eq.~(\ref{Equ:time2}) was used for $p[t_n]$ and the order of the finite sum and the integral was exchanged. The brackets $[\cdot]$ indicate the discrete nature of the argument, $t_n=nT_s$, and $\omega_m'=2\pi m \left(T_s N_g\right)^{-1}$.
The frequency resolution of the DFT spectrum is thus $\Delta\!f_{\mathrm{DFT}} = f_s N_g^{-1} = T_g^{-1}$. Due to the finite support of $g$, the sum in Eq.~(\ref{Equ:STFT1}) can be extended to $n\in \mathbb{Z}$ leading to the discrete time Fourier transform (DTFT, cf. \cite{Oppenheim1998}) of $g$ which is continuous in frequency and periodic with a period of $f_s$. 

Thus, Eq.~(\ref{Equ:STFT1}) can be written as:
\begin{align}\label{Equ:STFT3}
  DFT_g(p)[\omega'_m] =  \frac{ 1}{ 2 \pi}  \int\limits_{-\infty}^{\infty}  \phat\left(\omega\right) \sum\limits_{n=-\infty}^{\infty}{ \!\! \E^{\I t_n (\omega_m' - \omega)} g[t_n] }\ d\omega                                                                                                   = \frac{ 1}{ 2 \pi}  \sint  \phat\left(\omega\right) \mathcal{F}_{\text{DTFT}}(g)(\omega_m'-\omega) d\omega,
\end{align}
which is simply the convolution of the Fourier transform of $p$ and the DTFT of $g$, that allows the inclusion of the spectral leakage effect of the DFT into the transfer function.  
With this formulation, the windowed DFT of the sampled version of $p(t)$ can be calculated directly from the continuous frequency domain formulation 
without the need of an inverse transform to the time domain. 

The amount of spectral leakage depends on two factors: type and length of the window.  First, the higher $T_g$, the longer the window and the smaller the leakage. Second, the choice of window type is crucial for the reduction of spectral leakage. When a rectangular window of length $T_g$ is used ($g[t_n]=1,\, n = 0,\dots,N_g-1$), $\mathcal{F}_{\text{DTFT}}(g)(\omega_m'-\omega)$ is the Dirichlet kernel $D_{N_g}\left((\omega_m'-\omega)T_s\right)$ which resembles, loosely speaking, a periodized $\mathrm{sinc}$-function with period $2\pi$ that only slowly decays towards $\pm \pi$. For the Hanning window, e.g. the DTFT of $g[t_n]$ consists of the sum of three Dirichlet kernel functions which results in a faster decay of $\mathcal{F}_{\text{DTFT}}(g)$ and considerably reduced spectral leakage, which will be illustrated in the next section.  

Reintroducing the subscripts $\ell$ and $n$ and plugging Eq.~(\ref{Equ:UniIndP}) into Eq.~(\ref{Equ:STFT3}) leads to
\begin{align}\label{Equ:UniDFTt}\phat_{n\ell}[\omega_m'] &:= \text{DFT}_g(p)[\omega_m'] = \frac{1}{(2 \pi)^2} \dint  \shat_\ell(\omega - v_s k_x) \q{n\ell}{\sqrt{\omega^2/c^2-k^2_x}} \hat{g}(\omega_m'-\omega) \E^{\I k_x (x_{r,n} - x_{s,\ell})}  dk_x  d\omega,
\end{align}
which is the DFT of the pressure signal given at the $n$-th receiver resulting from the $\ell$-th moving source. Note that in contrast to Eq.~(\ref{Equ:UniIndP}),  $\phat$ now denotes the windowed DFT of $p[t_n]$ which is indicated using brackets $[\cdot]$. $\hat{g}$ on the other hand denotes the DTFT of $g[t_n]$, which is continuous in frequency, as indicated using parentheses.

\subsubsection{Single-frequency sources}
No assumptions about $\shat_\ell$ were used in Eq.~(\ref{Equ:UniDFTt}). A reasonable assumption to simplify the formulation is to use a harmonic ansatz to approximate the source signal $s_\ell(t)$  similar to, e.g. \cite{Gombots2021}. Thus, each source signal comprises a sum of complex exponentials
\begin{align}\label{Equ:CompExp}
s_\ell(t) = \sum\limits_{j=1}^{J} a_{\ell j} \E^{-\I \omega_j t},
\end{align}
 with frequencies $\omega_j=2\pi f_j$ and corresponding complex amplitudes $a_{\ell j}$ which are constant in time. 
 
 \begin{figure}[!ht]
\begin{center}
\includegraphics[trim=0cm 0cm 0cm 0cm, clip=true, width=0.8\textwidth]{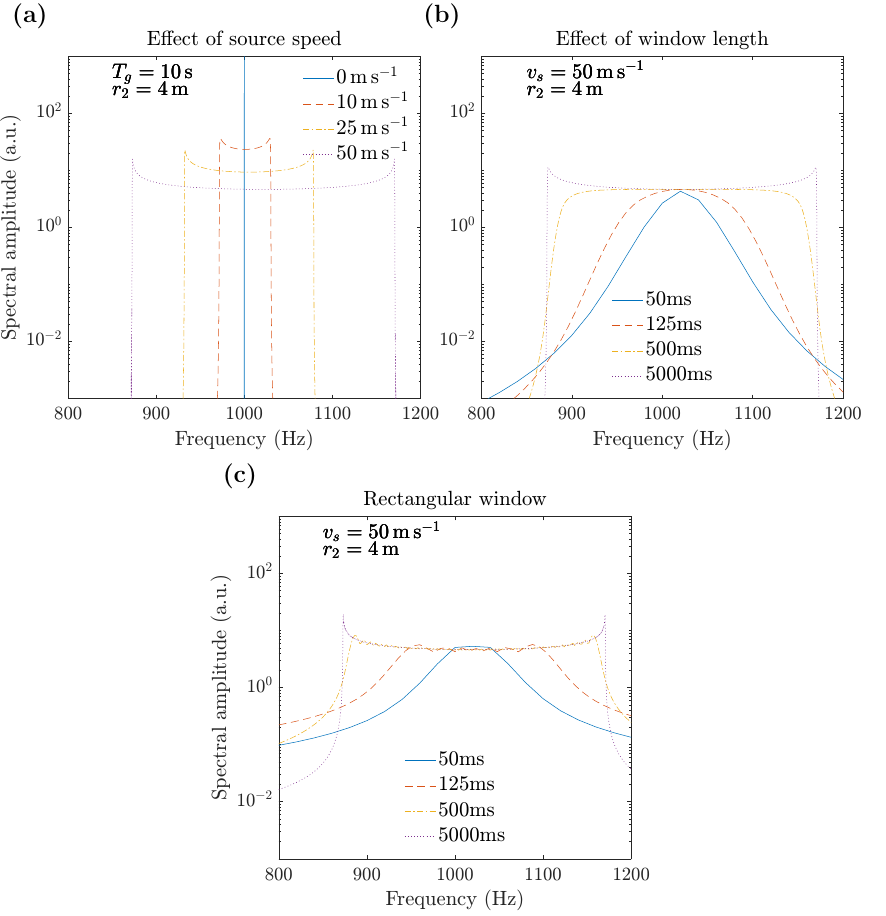}
\caption{DFT of the signal of a moving point source at a stationary receiver. The panels show the spectral amplitudes for a single-frequency source moving at different speeds $v_s$ (a), for different window lengths $T_g$ for a Hanning-window (b), and for a rectangular window (c). The distance between source and receiver plane was set to $r_2=4$\,m. }
\label{fig:motioneffect}
\end{center}
\end{figure}

 In the stationary case, each frequency $f_j$ can be treated separately, as long as the spectral leakage is adequately controlled for. If the sources are moving, each single frequency component leads to a range of observed frequencies  at the stationary receivers due to the Doppler effect. This spectral range depends on the speed of the source, which in turn has an effect on the DFT and the choice of all its parameters like sampling rate, window length, and window type (see Fig.~\ref{fig:motioneffect}(a) to (c) where a source with $\Fm=1000$\,Hz and a source-receiver distance in the $y$-$z$-plane of $r_2=4$\,m was used). 
Consequently, if the spectral ranges for different source components are overlapping, an individual treatment of the respective complex exponential is no longer possible and a joint treatment is necessary, which is beyond the scope of this work. 
Thus, the focus of the present work will be on single-frequency sources, i.e. a single complex exponential signal with frequency $\Om=2\pi f_0$. 
The source signal at the $\ell$-th source grid position $\bx_{s,\ell}$ is set to $s_{\ell}(t)=a_{\ell} \E^{-\I \Om t}$ with the Fourier spectrum $\shat_{\ell}(\omega)=a_{\ell} 2 \pi \delta(\omega-\Om)$ where $\delta(\cdot)$ denotes the Dirac delta distribution. 

Applying the spectrum to Eq.~(\ref{Equ:UniDFTt}) yields
\begin{align}
\nonumber \phat_{n\ell}[\omega_m'] &= \frac{a_\ell}{2 \pi} \dint \delta(\omega - v_s k_x-\Om)  \q{n\ell}{\sqrt{\omega^2/c^2-k^2_x}} \hat{g}(\omega_m'-\omega) \E^{\I k_x (x_{r,n} - x_{s,\ell})}  dk_x  d\omega \\
  \nonumber &=  \frac{a_\ell}{ 2 \pi |v_s|} \sint   \q{n\ell}{\sqrt{\omega^2/c^2-(\omega-\Om)^2/v_s^2}} \hat{g}(\omega_m'-\omega) \E^{\I (\omega - \Om)v_s^{-1} (x_{r,n} - x_{s,\ell})} d\omega \\
   \label{Equ:Monoxom}           &:= h_{n\ell}[\omega'_m]a_\ell
\end{align}
$\phat_{n\ell}$ represents the (noise-free) complex sound pressure amplitude at a discrete frequency $\omega_m'$ that would be ``measured'' at the $n$-th receiver for a moving single-frequency source of frequency $\Fm$ located at $\bx_{s,\ell} = (x_{s,\ell} + vt, y_{s,\ell}, z_{s,\ell})^\top$ when using a windowed DFT with the time window $g$. $h_{n\ell}$ is defined as the transfer function from the $\ell$-th source to the $n$-th receiver.  
\subsection{Numerical quadrature}
 A standard adaptive quadrature was applied in this manuscript using the function \texttt{quadgk} from MATLAB \cite{MATLAB} which is based on a Gauss-Kronrod quadrature. It has been shown previously~\cite{Kasess2016} that certain steps and special quadrature methods can be applied to efficiently solve the integral for the stationary case, which is of the form given in Eq.~(\ref{Equ:25D}). How this method can be applied to the current setting of a moving source is ongoing work.

A major difficulty of this inverse transformation lies in determining the integration limits which has a direct impact on the error but also on the computational effort. Every quadrature node\,/\,function evaluation implies the calculation of $\hat{q}_{n\ell}$, which, depending on the research question, may require a substantial amount of calculations.

\begin{figure}[!ht]
\begin{center}
\includegraphics[trim=0cm 0cm 0cm 0cm, clip=true, width=0.99\textwidth]{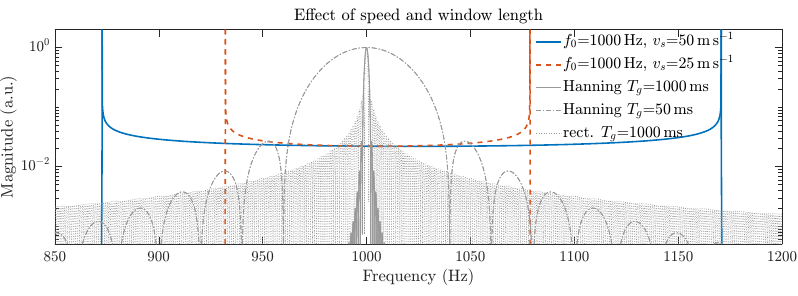}
\caption{Overview over the integrand. Shown is the magnitude of $\q{}{\cdot}$ in Eq.~(\ref{Equ:Monoxom}) for $\Fm=1000$\,Hz and source speeds of $v_s=25$\,m\,s$^{-1}$ (dashed orange) and $v_s=50$\,m\,s$^{-1}$ (solid blue). The thin gray lines show the Fourier transforms of a Hanning window: solid for $T_g=1000$\,ms, dash-dotted for $T_g=50$\,ms. The thin dotted gray line shows the Fourier transform of a rectangular window with $T_g=1000$\,ms. The Fourier transforms of the windows are normalized to their peak amplitude for easier comparison. } 
\label{fig:effectwindow}
\end{center}
\end{figure}

$\hat{q}_{n\ell}$ depends on Hankel functions which have a singularity at $0$, thus, $\hat{q}_{n\ell}$ has singularities at $\omega = c\,\Om(c \pm v_s)^{-1}$ which fulfill the relation $\omega^2/c^2 - (\omega - \omega_0)^2 v_s^{-2} = 0$. At these values, the integral in Eq.~(\ref{Equ:Monoxom}) has to be split. 
For $\omega < c\,\omega_0\left( c + v_s \right)^{-1}$ or $\omega > c\,\omega_0\left(c - v_s\right)^{-1}$,  Hankel functions, and hence the 2D solutions $\hat{q}_{n\ell}$ decay exponentially  (cf. \cite{Abramowitz9}, Eq.~9.2.3). Thus, it is sufficient to restrict the infinite integral in Eq.~(\ref{Equ:Monoxom}) to a finite interval which is slightly bigger than $[c\omega_0 \left ( c + v_s \right)^{-1},c\omega_0 \left(c - v_s\right)^{-1}]$. This can be seen, for example, in  Fig.~\ref{fig:effectwindow} which shows the magnitude of the solution of the free-field 2D solution $\q{}{\cdot}=H_0\left(\cdot\right)$ in Eq.~(\ref{Equ:Monoxom}) at $f_0 = 1000$\,Hz for two different source speeds: 50 m\,s$^{-1}$: solid blue line, 25 m\,s$^{-1}$: dashed orange line.
To illustrate the role of the window, Fig.~\ref{fig:effectwindow} also shows $\hat{g}$ for three different time windows (gray thin lines) for a single source-receiver combination with $r_2=4$\,m distance and $\Fm$=1000\,Hz. The solid and dash-dotted grey lines show the spectra of a Hanning window for different window durations. For short time windows, the Fourier transform of the window covers a wide frequency range and is only slowly decaying, thus the computational effort for the quadrature is higher. For reasons of comparison, the DTFT of a rectangular window of 1000\,ms length (dotted thin gray line) is also shown, and the larger spectral leakage can be easily seen when compared to the Hanning window of the same length. The consequence at the measurement position can be seen when comparing Fig.~\ref{fig:motioneffect}(b) and (c). 
   
\subsection{Source localization using the transfer function}

Up to now, all derivations used radial frequencies ($\Ompm{m}$, $\Om$) to avoid a cluttered notation due to factors of $2\pi$. From now on, for practical reasons the notation will be switched to regular frequencies ($\Fmpm{m}$, $\Fm$).
With the transfer function $h_{n\ell}$ defined in Eq.~(\ref{Equ:Monoxom}), the total spectral complex amplitude at the receiver $n$ caused by all possible sources is given as: 
\begin{align}\label{Equ:Transfer}
\phat_n[\Fmpm{m}]  &= \sum\limits_{\ell=1}^{L} a_\ell h_{n\ell}[\Fmpm{m}],
\end{align}
or, if the pressures at all microphones is collected into one vector $\mathbf{\phat} = (\phat_1, \dots, \phat_N)^\top$ and the unknown source strengths into $\mathbf{a} = (a_1,\dots,a_L)^\top$,
 \begin{align}\label{Equ:TransferMat}
\mathbf{\phat}[\Fmpm{m}]  &= \mathbf{H}[\Fmpm{m}] \mathbf{a}.
\end{align}
As the difference between calculated values $\hat{\mathbf{p}}$ and measured sound pressures $\tilde{\mathbf{p}}$ should be as small as possible, the unknown  amplitude  vector ${\bf a}$ can be calculated by solving the least squares problem 
\begin{equation}\label{Equ:Optim}
  \underset{\bf{a} \in \mathbb{C} }{\mathrm{min}} \lVert{\mathbf H}[\Fmpm{m}] {\bf a} - \tilde{{\mathbf p}} [\Fmpm{m}] \rVert_2^2.
\end{equation}
This is the basic optimization problem for a single frequency $f'_m$, which is typically strongly underdetermined, as the number of microphones $N$ is much smaller than the number of potential sources $L$. In contrast to a stationary setting where only one specific frequency bin $\Fmpm{m}\approx\Fm$ can be used, the Doppler shift allows for the use of a spectral range $f'_m \in [f_-,f_+]:= [\Fm (1 + v_s/c)^{-1} , \Fm (1 - v_s/c)^{-1}]$ in Eq.~(\ref{Equ:Optim}).

From this  range it is possible to select $M$  discrete  frequency bins, where $M$ is (up to some uncertainty at the frequency limits) restricted to $M \leq 1 + (f_+ \! - \! f_-)\, \Delta\!f_{\mathrm{DFT}}^{-1}$. For a clearer notation, these frequency bins are collected in a set $\Omega=\left\{ \Fmpm{m}\, |\, f_+\!\leq\! \Fmpm{m}\! \leq\! f_- \right\}$ of cardinality $|\Omega|=M$. 
This approach has the advantage that instead of a highly underdetermined $N\times L$ system with $N \ll L$, a more stable $(N\cdot M) \times L$ system can be used for the least square approach. A suitable spectral range depends on many factors (cf. Fig.~\ref{fig:motioneffect}(a) and (b)), further restricting $f_{\mp}$. Setting the limits too wide may lead to some observations being more susceptible to corruption by background noise.
 How to determine a suitable set with regard to the size $M$ and choice of frequencies will be investigated in Section~\ref{Sec:Results}.

The case of the same single set $\Omega$ used for all microphones may be generalized to $N$ different sets $\Omega_n$ such that a different set of frequencies $f'_{m(n)}$ is selected for each of the $N$ microphone measurements. 
In addition to increasing the number of observations by choosing $M>1$, regularization is an important topic when solving Eq.~(\ref{Equ:Optim}). In this work, a standard Tikhonov regularization procedure combined with an L-curve approach is used to assess the effect of different parameters of the numerical model, e.g. source signal frequency, window length, speed, and choice of receiver frequencies. The resulting maps ${\bf a}$ show the spread of the point source, similar to a point-spread-function, and allow conclusions about how to control this spread (see Section~\ref{Sec:Results}). 

After all measurement data have been transformed using a windowed DFT,  $N\!\cdot\!M$ measurement points in the discrete frequency domain are selected, comprising $N$ vectors $\tilde{\bf{p}}_n[\Omega_{n}]$ of dimension $M$ containing data according to the frequencies defined in the respective set $\Omega_n$.  Applying the standard Tikhonov-regularization (also called ridge regression) minimizes the following expression:
\begin{align}
\label{Equ:Tikh}
  \underset{\bf{a} \in \mathbb{C} }{\mathrm{min}} \sum\limits_{n=1}^{N}{ \left\lVert \tilde{\bf{p}} _n[\Omega_n] -  \sum\limits_{\ell=1}^{L} a_\ell 
  h_{n\ell}[\Omega_n] \right\rVert_2^2} +  \lambda \sum\limits_{\ell=1}^{L} \left| a_\ell \right| ^2
\end{align}
or more compactly
\begin{align}
\label{Equ:TikhM}
\underset{\bf{a} \in \mathbb{C} }{\mathrm{min}}  \lVert{   \tilde{\mathbf p}_\Omega - {\mathbf H_\Omega} {\mathbf a} }\rVert_2^2 + \lambda \lVert{\mathbf a}\rVert_2^2,
\end{align}
where ${\mathbf H_\Omega}$ is of dimension $(M\cdot N)\times L$. 
Finding the optimal regularization parameter $\lambda$ is done using the L-curve approach \cite{Hansen1993}, as implemented in the MATLAB library \textit{Regularization Tools} \cite{Hansen2007}.

\subsection{Validation procedure}
\begin{figure}[!ht]
\begin{center}
\includegraphics[trim=0cm 0cm 0cm 0cm, clip=true, width=0.45\textwidth]{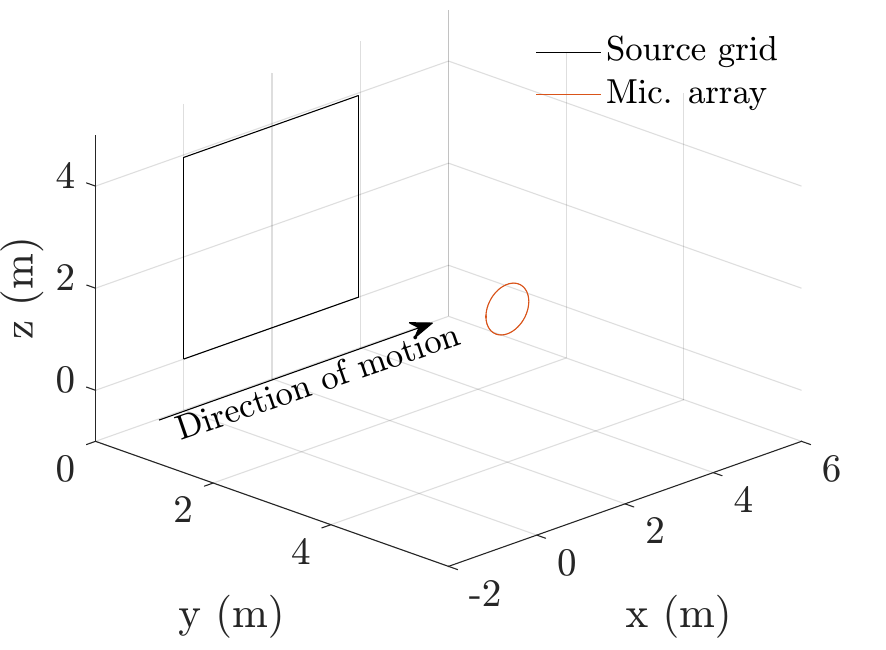}
\caption{Area covered by the source grid (black rectangle) at time $t = 0$\,s and the area covered by the stationary microphone array (orange circle) which was used for the test cases. 
}
\label{fig:grid}
\end{center}
\end{figure}

The following setup was created to validate the inverse approach. First, potential sources are assumed to be located on a 4\,m$\times$4\,m grid in the $x$-$z$-plane which is moving along the $x$-axis. This grid is discretized using an isotropic spacing of 0.05\,m resulting in an 80$\times$80 grid of source points. The receiver positions were defined by a  microphone array of 112 microphones with a geometry based on spirals and a diameter of approximately 1\,m. The array was placed parallel to the moving source grid at distances of 4, 6, and 8\,m in the $y$-direction. 
Fig.~\ref{fig:grid} shows the arrangement for the simulated data, where the source area is depicted at $t=0$\,s. The algorithm's performance is evaluated in a free 3D-space as well as above a fully reflecting half-plane located $z=-1$\,m. 

The sound pressure at the microphones was created using simulated data of a uniformly moving single-frequency point source with frequencies $f_0$ = 250, 500, 1000, and 2000\,Hz, and source speeds of $v_s$ = 1, 10, 25, and 50\,m\,s$^{-1}$. The source itself was located at the center of the source grid at $(x_0,y_0,z_0) = (2,0,2)$ at time $t = 0$\,s. 
A simple time formulation was used to generate the test signal (cf., \cite{DeHoop2005,Duhamel1996}). For the experiments including a reflecting half-plane (see Section~\ref{Sec:Ground}), the source is mirrored at the plane $z = -1$ and added to the original source. 
All signals were generated using a sampling frequency of $f_s$=10\,kHz, which is sufficiently high to accurately resolve the Doppler shifted frequencies for all $f_0$ and $v_s$ considered.

One important aspect of the numerical experiments is to evaluate different choices for $\Omega_n$ and the window function $g$ with respect to the performance as well as the gain in resolution by computing the convolution. For the time window $g$ Hanning windows of 7 different lengths $T_g$ ranging from 50\,ms to 5000\,ms were considered for the test signals as well as the transfer functions. The windows were centered around $t=0$\,s. 
To calculate the integral in Eq.~(\ref{Equ:Monoxom}), the integration limits were set such that $\hat{g}(\omega_m'-\omega)$ in Eq.~(\ref{Equ:Monoxom}) was more than 80\,dB below the central peak $\hat{g}(0)$. 

As the L-curve approach is known to occasionally lead to multiple corners and thus no clear optimal regularization value, a small amount of noise was added as suggested in \cite{Johnston2000} where they showed that poor L-curve shapes may occur due to ``correlated noise'', e.g. positioning errors. The noise was scaled to achieve a signal-to-noise ratio (SNR) of 80\,dB compared to the peak signal value.  In the results presented here, the quadrature error in the frequency domain compared to the time domain approach used to generate the microphone data may be a reason for this effect. Overall, the small amount of noise lead to a much more stable regularization.

\section{Results}\label{Sec:Results}

\subsection{Selection of observation frequencies $\Omega_n$}
\begin{figure}
\begin{center}
\includegraphics[trim=0cm 0cm 0cm 0cm, clip=true, width=\textwidth]{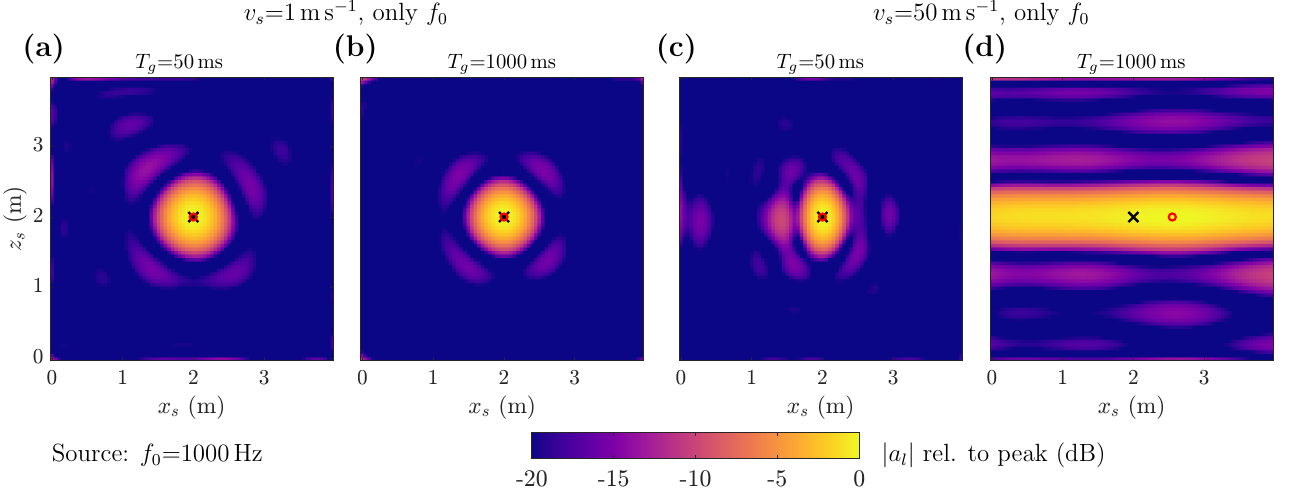}
\caption{Source maps based on a single DFT bin $\Omega_n=\left\{f_0\right\}$ for each microphone. Shown are the amplitudes of the regularized solutions for a source with $f_0=1000$\,Hz moving with two velocities: $v_s=1$\,m\,s$^{-1}$ in panels (a) and (b), $v_s=50$\,m\,s$^{-1}$) in panels (c) and (d). A short window ($T_g=50$\,ms) was used in panels (a) and (c),  a long window  with $T_g=1000$\,ms was used in panels (b) and (d). 
  The black x denotes the true source position, the red circle the position of the detected maximum. The dynamic range is set to 20\,dB and the maps are normalized to the respective peak amplitude.} 
\label{fig:singlefreq}
\end{center}
\end{figure}

It was already mentioned in the previous sections that for a single-frequency source with frequency $\Fm$ the measured signal at a stationary observer contains a frequency spectrum due to the Doppler effect. Important factors determining the width of this spectrum  are the source frequency $\Fm$, the speed of the source $v_s$, the length $T_g$ (cf. Fig.~\ref{fig:motioneffect}) as well as the temporal position of the time window, and the distance between source and receiver. The choice of the sampling set(s) $\Omega_n$ of measurement frequencies used for the inversion of the transfer matrix is an important issue. In order to indicate the discrete nature of the DFT spectrum and the associated bandwidth $\Delta\! f_{\mathrm{DFT}}$ in the spectrum, the term DFT bin will be used for brevity when data at a specific frequency of the DFT spectrum is meant. 

\subsubsection{Single frequency bin}
In a first setting, the single DFT bin closest to the source frequency was used for each microphone, i.e. $\Omega_n=\left\{f_0\right\}, \forall n = 1,\dots,N$ (up to a potential frequency mismatch due to the DFT bin spacing). 
This is closely related to the stationary case, where only the signal frequency is available. For the setting used in the evaluation, the corresponding transfer matrix ${\mathbf H_\Omega}$  is of dimension 112$\times$6400 (number of receivers times number of potential sources). The entries of the matrix ${\mathbf H_\Omega}$  were calculated using Eq.~(\ref{Equ:Monoxom}). 
Considering an almost stationary source ($v_s = 1$\,m\,s$^{-1}$), the regularized inverse leads to the source maps ${\bf a}$ shown in Fig.~\ref{fig:singlefreq}(a) and (b). Similar to conventional beamforming, the detected source position is blurred. However, the peak (red circle) is located at the true source position (black $\times$). The blurring of the source is roughly the same in $x$ and $z$.

The window length seems to have only a minor effect on the result, although when going to even longer windows (5000\,ms, not shown), the main lobe is more concentrated around the true source position but also two sidelobes with roughly -9\,dB appear along the $x$-direction. Note that all source maps are normalized to the peak amplitude since only the spatial blurring was of interest in the current work. 

For the fast moving situation (Fig.~\ref{fig:singlefreq}(c) and (d)), a number of effects occur. While the localization in the vertical direction ($z$) is essentially unaffected, a smearing occurs along the direction of motion for longer time windows (Fig.~\ref{fig:singlefreq}(d), 1000\,ms). For the given speed, the source moves 50\,m during the period of 1000\,ms, while it only moves 2.5\,m during 50\,ms. Thus, the smearing along $x$ occurring for longer analysis windows seems intuitively clear. On the other hand, the motion of the source as well as the spectral leakage caused by the DFT is, in theory, already considered in the transfer function. Apparently, the information contained at the same frequency across all microphones is not sufficient to lead to a reasonable estimate for the source map. 
In addition to the blurring, the peak (red circle) in the source strength map is also considerably displaced from the true source position (black x). 
The reason for this blurring along $x$ for increasing $T_g$ is that the elements in the transfer matrix  (Eq.~\ref{Equ:Monoxom}) converge towards the 2D solution multiplied by a phase factor depending on $x$ and the chosen bin. Thus, when the same single frequency bin is used across all receivers in connection with a long window, shifts in the $x$-position of a source can be approximated by a phase shift in its complex amplitude leading to an ambiguity. The formal details for the limiting case $T_g \rightarrow \infty$, where the relation between source position and phase is exact, are given in \ref{App:SingleCase}.

As a consequence, restricting the inverse scheme to the same single DFT bin for all microphones does not yield satisfying results. As illustrated in \ref{App:SingleCase} the increased blurring along $x$ is found in connection with long windows and is not depending on the actual bin used. Thus, in a next step a set of DFT bins arranged around the source frequencies is considered for the inversion process taking advantage of the spectral spread induced by the motion of the source. This has the effect of more observations and thus a less heavily underdetermined system of equations. 
\subsubsection{Regularly spaced frequency bins}
\label{Sec:RegBins}
Before addressing the actual choice of frequencies used for setting up the transfer matrix, the available frequency range caused by the Doppler effect and the window was determined. The equation for the Doppler shift of a moving source yields a maximum range given by $f_0 (1\pm v_s/c)^{-1}$. For example, assuming a source speed of $v_s  = 50$\,m\,s$^{-1}$, a frequency of $\Fm=1000$\,Hz, and the speed of sound to be $c=343$\,m\,s$^{-1}$ leads to a spectral range between approximately 873\,Hz and 1170\,Hz, if a long enough temporal segment is considered.

To allow for a better comparison of the source maps, the same frequency range was used for all analysis window lengths investigated (Hanning windows with $T_g = 50$\,ms to $T_g = 5000$\,ms) as shorter temporal segments for the DFT lead to a reduced observed spectral spread. The range of frequencies for the upcoming analyses was restricted to $f_-=920$\,Hz to $f_+=1120$\,Hz. Close to these frequency bins, a decay in amplitude of roughly 40\,dB for $T_g = 50$\,ms (c.f. Fig.~\ref{fig:motioneffect}(b)) can be observed. For longer windows, the decrease in amplitude at these frequencies is considerably smaller.

\begin{figure}
\begin{center}
\includegraphics[trim=0cm 0cm 0cm 0cm, clip=true, width=\textwidth]{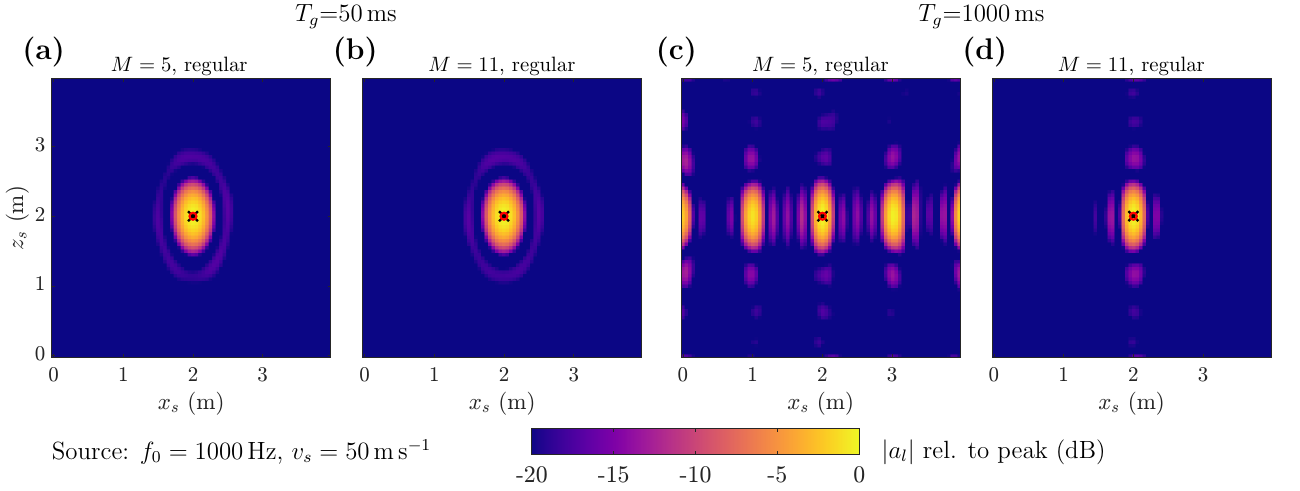}
\caption{Source maps based on the same set of regularly spaced frequencies for each microphone. Shown are the amplitudes of the regularized solutions for $\Fm=1000$\,Hz and $v_s=50$\,m\,s$^{-1}$ using a short window ($T_g=50$\,ms) in panels (a) and (b), and a long window ($T_g=1000$\,ms) in panels (c) and (d). Panels (a) and (c) show the case of $M=5$; panels (b) and (d) show the case of $M=11$ regularly spaced frequency bins. The black x denotes the true source position, the red circle the position of the detected maximum. The dynamic range is set to 20\,dB and the maps are normalized to the respective peak amplitude.} 
\label{fig:equifreq}
\end{center}
\end{figure}

Based on the values for $f_\pm$, a suitable way to choose DFT bins from the available frequency range must be defined. The first scheme will employ the same set $\Omega_n$ of equally spaced DFT bins covering the predefined frequency range for each microphone. Fig.~\ref{fig:equifreq}(a) and (c) show the results for selecting $M = 5$ and Fig.~\ref{fig:equifreq}(b) and (d) for $M = 11$ (nearly) equally spaced frequency bins leading to matrix sizes of 560$\times$6400 and 1232$\times$6400, respectively. 
Note that a regular spacing is not always possible depending on the window length, frequency limits, and $f_0$. For $T_g=50$\,ms, for example, the spacing of the DFT bins is given by $\Delta\! f_{\mathrm{DFT}}=20$\,Hz and the restricted spectrum ranging from 920\,Hz to 1120\,Hz contains 11 bins as the limits also coincide with the bin frequencies. With this setting, a regular spacing is therefore only achieved for $M = 2, 3,6,$ or $11$.  For longer windows, deviations from regular grids occur less often because $\Delta\! f_{\mathrm{DFT}}$ is smaller.

A comparison of Figs.~\ref{fig:equifreq}(a) and (b) with Fig.~\ref{fig:singlefreq}(c) shows much smaller sidelobes for the case of a 50\,ms Hanning window if more then one DFT bin is used. For the long Hanning window with length $T_g=1000$\,ms (panels (c) and (d) in Fig.~\ref{fig:equifreq}), the situation is more complex. Note that for $T_g = 1000$\,ms, the DFT bin spacing is given by $\Delta\! f_{\mathrm{DFT}} = 1$\,Hz.  
For $M=11$ bins (panel (d)), the $x$-resolution is massively improved compared to the case where only one single bin was used (cf. Fig.~\ref{fig:singlefreq}(d)). If $M = 5$, the main lobe is better localized around the true position but periodic sidelobes appear. 

The periodicity of 1\,m for the biggest lobes is a consequence of the frequency spacing chosen. Considering the range spanning 200\,Hz, $M = 5$ bins are equally spaced lying 50\,Hz apart. This regular spacing of 50\,Hz is equivalent to a period length of 20\,ms which, at $v_s = 50$\,m\,s$^{-1}$, is equivalent to a traveled distance of 1\,m. 

By the same rule, results from an inversion using 4 bins exhibit a periodicity of 0.75\,m (not shown here). Interestingly, there also is a small, but noticeable damping acting on these ``sidelobes". 
The seemingly shorter period of roughly 0.25\,m results most likely from a periodic superposition of the sidelobes also occurring e.g. in Fig.~\ref{fig:equifreq}(d). 

\begin{figure}[!ht]
\begin{center}
\includegraphics[trim=0cm 0cm 0cm 0cm, clip=true, width=0.9\textwidth]{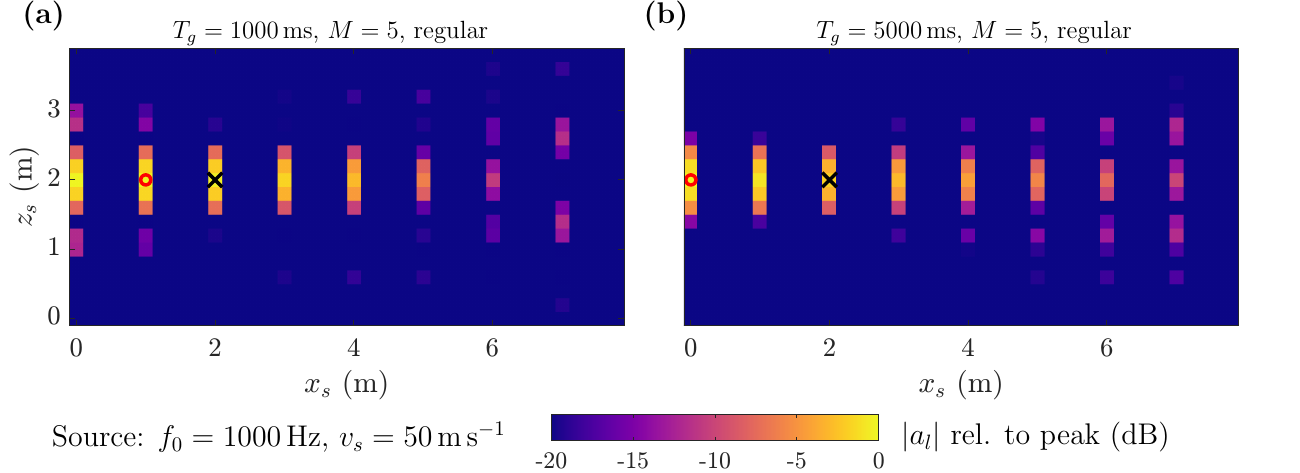}
\caption{Effect of window length on the source map using the same set of $M=5$ regularly spaced frequencies for each microphone. Shown are the amplitudes of the regularized solutions for $\Fm=1000$\,Hz when using $T_g=1000$\,ms (a) and $T_g=5000$\,ms (b). The black x denotes the true source position, the red circle the position of the detected maximum. The dynamic range is set to 20\,dB and the maps are normalized to the respective peak amplitude.} 
\label{fig:equifreqdetail}
\end{center}
\end{figure}

To investigate the periodicity and the slight damping in $x$ in more detail, a larger source grid ranging from 0 to 8\,m with a smaller source grid spacing of 0.2\,m (isotropic) was used. In this setting it is clearly visible that there is a mismatch between the position of the maximum amplitude in the source map and the true position. Furthermore, the decay with respect to $x$ is now better visible in the source maps shown in Fig.~\ref{fig:equifreqdetail}, and comparing the 1000\,ms (left, a) and the 5000\,ms (right, b) window, the periodic sidelobes decay faster for the shorter window. This effect is related to the increased spectral leakage caused by short windows. Briefly, in the extreme case where the window length  $T_g\rightarrow \infty$ no leakage occurs. In this case, the source map will become periodic up to a phase factor as discussed in detail in \ref{App:Regular}.
For sufficiently short windows, no periodicity is observed (Fig.~\ref{fig:equifreq}(a)). In addition to the high spectral leakage, short windows lead to a coarser frequency resolution in the DFT. This may lead to a situation where the DFT bin spacing does not exactly align with the chosen frequency spacing leading to an irregularly spaced set of frequency bins which results in a further deviation from the periodic limiting case analyzed in \ref{App:Regular}.
For example, using a window length of $T_g = 50$\,ms the DFT bin spacing is $\Delta\! f_{\mathrm{DFT}}=20$\,Hz. Using a frequency range of 200\,Hz and $M = 5$ bins results in a required regular spacing of the selected DFT bins of 50\,Hz. Therefore, no regular frequency grid spacing could be achieved with a DFT bin spacing of $20$\,Hz. 

\subsubsection{Randomly selected bins}\label{Sec:Random}
To avoid the periodicity problem, a different approach is suggested which uses a set of randomly selected frequency bins from the available frequency range. In detail, for each of the 112 microphones, $M$  frequency bins are randomly chosen within the available frequency range. Thus, contrary to the approach using a regular spacing, DFT bins may vary across microphones. 
However, as already described in the case of regularly spaced bins, the number of available different frequency bins to choose from depends not only on the predefined frequency range but also on frequency resolution which is the inverse of the chosen window length.   

\begin{figure}
\begin{center}
\includegraphics[trim=0cm 0cm 0cm 0cm, clip=true, width=\textwidth]{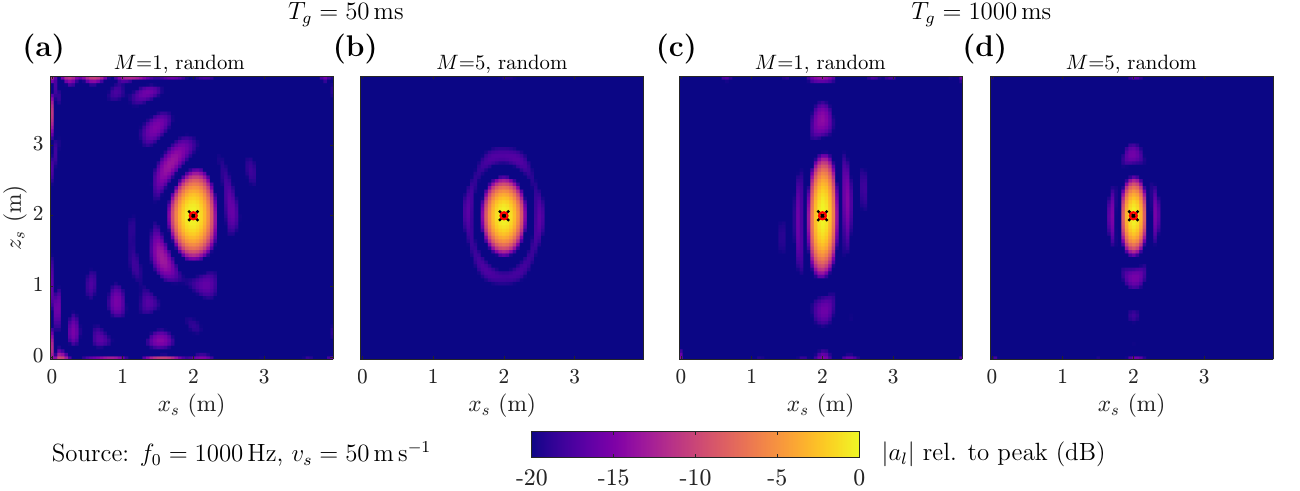}
\caption{Source maps for different sets of random frequencies for each microphone. Shown are the amplitudes of the regularized solutions for $\Fm=1000$\,Hz 
and $v_s=50$\,m\,s$^{-1}$ using a short window ($T_g=50$\,ms) in panels (a) and (b) and a long window ($T_g=1000$\,ms) in panels (c) and (d). Panels (a) and (c) show the case of $M=1$ randomly assigned frequency bin per microphone, panels (b) and (d) the case of $M=5$ randomly assigned frequency bins per microphone. The black x denotes the true source position, the red circle the position of the detected maximum. The dynamic range is set to 20\,dB and the maps are normalized to the respective peak amplitude.} 
\label{fig:randomfreq}
\end{center}
\end{figure}

Keeping the same values for $f_\pm$ as before, Fig.~\ref{fig:randomfreq} illustrates the effect of using random frequencies. Clearly, the periodic effect has disappeared while the resolution in the $x$-direction is greatly improved compared to the single frequency bin case, in particular for longer windows. For short time windows (panels (a) and (b)), there is no clear difference between using the same number of observations for a regular spacing (Fig.~\ref{fig:equifreq}(a)) and a random selection (Fig.~\ref{fig:randomfreq}(b)). First, for the short window there was no periodicity present in the regular case and, secondly, the number of available bins is relatively small. For a range of 200\,Hz and a window length of 50\,ms this amounts to 11 bins per microphone and thus the randomly chosen set of frequencies will be similar to the regular choice. 
In contrast, for $T_g=1000$\,ms (panels (c) and (d)), 201 frequency bins are available per microphone and thus the set will be sufficiently random to omit periodic effects that occur for a regular spacing in this case. On the downside, for $M=1$ and the longer window, the vertical spreading increases compared to the previous cases using the signal frequency or regularly spaced bins. Setting $M=5$ leads to a similar vertical spread compared to 11 equidistantly spaced DFT bins.

\subsection{Effect of source frequency $f_0$}
Already in the stationary case, the frequency $\Fm$ of the source is an important factor affecting the  localization performance of microphone arrays. 
The Doppler effect caused by a moving source additionally results in a spectral spreading of the observed frequency range that scales with $\Fm$. Lower $\Fm$ results in fewer available frequency bins that can actually be used for the inverse source localization, i.e. the choice of $M$ also depends on $\Fm$.

In this section, the effect of the source frequency is investigated using values of $\Fm$ = 250, 500, 1000, and 2000\,Hz. As both, $\Fm$ and $T_g$, affect $M$, the effect of $\Fm$ cannot be investigated independently of $T_g$ and thus a range of values for $T_g$ from 50\,ms to 5000\,ms is considered. 
The same geometries and grids as in the previous section were used. A moving source with velocity $v_s= 50$\,m\,s$^{-1}$ was positioned at $(2,0,2)$ at $t=0$\,s. Like in the previous section, the frequency range was set according to the speed using the same factor for all frequencies: $f_-=0.92\Fm$ and $f_+=1.12\Fm$.
The localization performance for different $f_0$ was examined for varying window lengths and different values for $M$ ranging from 1 to 5. Note that in the case of $f_0=250$\,Hz and $T_g=50$\,ms only 3 DFT bins are available for the given frequency range. Thus, the maximum value for $M$ was limited to 3 for this case.
The numerical experiments in the previous section were only based on one single set of random frequencies. In order to evaluate the variation of the results in dependence of the random realization, 10 different random sets of $\Omega_n$ frequencies were generated for each case investigated.   
As in the previous section, the analysis is based on the Tikhonov-regularized solution for the source grid.

\begin{figure}[!ht]
\begin{center}
\includegraphics[trim=0cm 0cm 0cm 0cm, clip=true, width=0.99\textwidth]{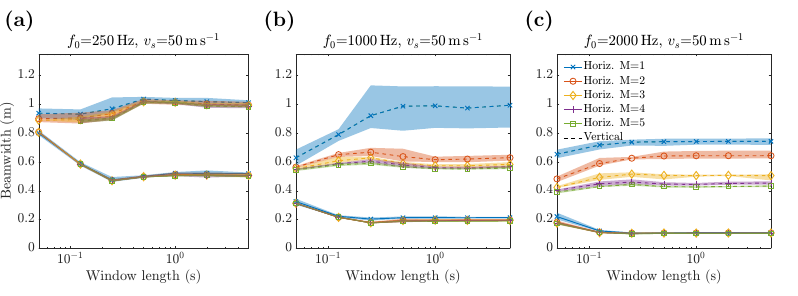}
\caption{Beamwidth as a function of frequency, window length, and number of bins. Shown are the mean horizontal beamwidth (solid colored lines and symbols) and the mean vertical beamwidth (dashed colored lines and symbols) for the regularized solution and a threshold of 3\,dB. Different values of $\Fm$ are shown in panels (a) to (c): 250, 1000, and 2000\,Hz. The source speed was set to $v_s=50$\,m\,s$^{-1}$. Colors and symbols denote different numbers of bins $M$.} 
\label{fig:effectfreq}
\end{center}
\end{figure}

To better illustrate the effect of the different parameters on the source localization, the -3\,dB-contour around the true position (which deviated at most one grid point from the detected main peak for all cases) was determined for each resulting source map using the MATLAB-function \texttt{contourc}. The horizontal ($x$) and vertical ($z$) extension of the contour around the true position were determined and plotted in  Fig.~\ref{fig:effectfreq} as a function of the analysis time window length.
In correspondence to beamforming, these two quantities will be referred to as horizontal (denoted by solid lines) and vertical beamwidth (denoted by dashed lines). Each panel in Fig.~\ref{fig:effectfreq} depicts the mean beamwidths across 10 random sets for frequencies $\Fm$ = 250, 1000, and 2000\,Hz, respectively.  The shaded areas depict the maximum and minimum beamwidth across the 10 random sets. 
Colors and symbols distinguish different values for $M$. For conciseness, the results for $\Fm=500$\,Hz which lie between the results for 250\,Hz and 1000\,Hz are not shown here.

As expected, the average resolution of the moving point source improves with increasing frequencies, i.e. the horizontal and vertical beamwidths decrease for increasing frequencies. There are, however, differences in the behavior of the horizontal and vertical beamwidth with respect to $T_g$ and $M$.

While the horizontal beamwidth is clearly reduced with increasing window duration, in particular for lower frequencies there is virtually no dependency on the number of frequency bins $M$. 

The vertical beamwidth, on the other hand, is very large in comparison and no reduction or even an increase of the beamwidth is achieved when increasing the window length.  Using an increasing number of frequency bins per microphone gradually improves the resolution in the vertical direction, at least for higher frequencies. For 250\,Hz (Fig.~\ref{fig:effectfreq}(a)), the number of bins ($M$) has no relevant effect on the beamwidth, which is probably a result of the reduced frequency range at low $\Fm$ as there are simply not many different frequency bins to chose from. For 500\,Hz (not shown), the reduction in horizontal beamwidth is also relatively small in the range of around 0.1\,m and below. 

Overall, Fig.~\ref{fig:effectfreq} shows that  the difference in beamwidth between $M=4$ and $M=5$ is relatively small, and there seems no point in going beyond $M=5$. A test calculation using $M=10$ did not yield any improvements compared to $M = 5$. 
Concerning the variation of the estimates, it is clearly shown that for $M=1$ the vertical beamwidth strongly varies with selected observation frequency bins $\Omega$ while for $M\!\!>1$  the localization performance is almost independent of the actual set of frequencies. The reason for the very high variability and deviation for $M=1$ and 1000\,Hz is, however, unclear. 
 
\subsection{Effect of source speed and distance between source and receiver}
\begin{figure}[!ht]
\begin{center}
\includegraphics[trim=0cm 0cm 0cm 0cm, clip=true, width=0.99\textwidth]{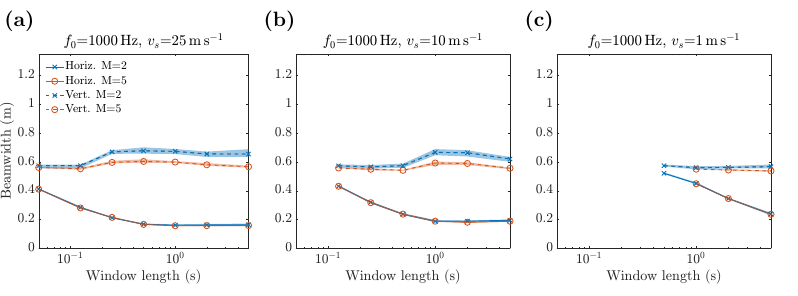}
\caption{Beamwidth as a function of speed, window length, and number of bins. Shown are the mean horizontal beamwidth (solid colored lines and symbols) and the mean vertical beamwidth (dashed colored lines and symbols) for the regularized solution and a threshold of 3\,dB. Different values of $v_s$ are shown in the panels (a) to (c): 50, 25, and 1\,m\,s$^{-1}$. The source frequency was set to $\Fm=1000$\,Hz. Colors and symbols denote the number of bins: blue x for $M=2$ and orange circles for $M=5$.}  
\label{fig:effectspeed}
\end{center}
\end{figure}

The second parameter affecting the Doppler effect is the source speed $v_s$. 
Lowering the speed from 50\,m\,s$^{-1}$ to 25, 10, or even 1\,m\,s$^{-1}$ leads to the results shown in Fig.~\ref{fig:effectspeed}.  Setting the frequency ranges in a similar manner as in the case of 50\,m\,s$^{-1}$, i.e. using $T_g=50$\,ms  as a reference and determining the closest bin to a decay of 40\,dB from the spectral peak was not possible, as the limits were beyond the maximum Doppler shift. Thus the bin frequencies closest to this criterion but within the maximum Doppler shift were chosen. 
In each panel, the beamwidths are shown for a different velocity and different number of DFT bins $M$ as a function of the window length $T_g$. In all panels the source frequency was set to $\Fm = 1000$\,Hz. For certain combinations with low source velocity (mainly $v_s=$1\,m\,s$^{-1}$) and short windows, the frequency range caused by the Doppler shift is too small for the DFT sampling to choose sufficiently many DFT bins for $M>1$. Hence, no results can be shown for these cases.

The beamwidth in $x$ decreases with increasing window length up to a certain point, but the number of used frequency bins does not affect the result very much. However, the number of DFT bins per microphone has some minor influence on the beamwidth in $z$. Using  $M=5$ leads to estimates similar across all values for $v_s$ (Fig.~\ref{fig:effectspeed} and Fig.~\ref{fig:effectfreq}(b)). 

Interestingly, the effect of the window length on the horizontal beamwidth depends on $v_s$, the decline of the curves shifting towards higher window lengths as the speed $v_s$ goes down and the final value is similar across different speeds. Although the exact relation was not determined, the data suggests that the shift is inversely proportional to the decrease in speed, i.e. halving the speed leads to the curve being ``shifted'' to double the window lengths. For example, the minimum window length for which the horizontal beamwidth does not decrease anymore is around 250\,ms for $v_s=50$\,m\,s$^{-1}$ as shown in Fig.~\ref{fig:effectfreq}(b), 500\,ms for $v_s =25$\,m\,s$^{-1}$, and around 1000\,ms for $v_s=10$\,m\,s$^{-1}$. For $v_s=1$\,m\,s$^{-1}$, the point of leveling-off cannot be determined. This effect seems reasonable considering that signal portions containing high energy will become longer by essentially the same factor.  Thus, a longer window will be better suited to cover the relevant signal portion. The variability of the beamwidth (shaded areas) is very low.
 
\begin{figure}[!ht]
\begin{center}
\includegraphics[trim=0cm 0cm 0cm 0cm, clip=true, width=0.66\textwidth]{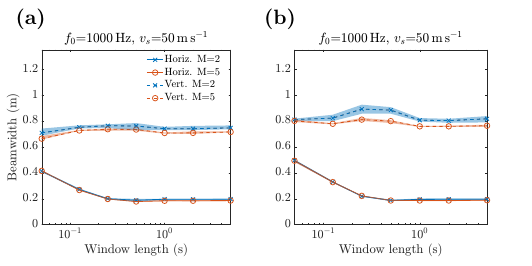}
\caption{Beamwidth as a function of source-receiver distance in $y$, window length, and number of bins. Shown are the mean horizontal beamwidth (solid colored lines and symbols) and the mean vertical beamwidth (dashed colored lines and symbols) for the regularized solution and a threshold of 3\,dB. Different values of the distance between sources and array are shown in the panels: 6\,m (a) and 8\,m (b). The source speed was set to $v_s=50$\,m\,s$^{-1}$ and the source frequency to $\Fm=1000$\,Hz. Colors and symbols denote the number of bins: blue x for $M=2$ and orange circles for $M=5$.} 
\label{fig:effectdist}
\end{center}
\end{figure}

When the distance between the source plane and the receiver plane is increased to 6\,m (Fig.~\ref{fig:effectdist}(a)) or to 8\,m (Fig.~\ref{fig:effectdist}(b)), the vertical beamwidth increases, i.e. the source map becomes less localized (cf. Fig.~\ref{fig:effectfreq}(b) for 4\,m distance) as is to be expected. 
In contrast, the much smaller horizontal beamwidth is essentially unaffected by the distance in $y$. 
 
\subsection{Effect of the window function}
As shown in Sec.~\ref{sec:meth}, employing a DFT leads to a convolution of the 2.5D solution with the Fourier transform of the time window which needs to be evaluated. In general, this is a costly operation, as it requires the calculation of many 2D solutions. The length and type of the window $g$ affects the integration limits and thus has a huge influence on the computational effort of the numerical convolution (cf. Fig.~\ref{fig:effectwindow}). 

\begin{figure}[!ht]
\begin{center}
\includegraphics[trim=0cm 0cm 0cm 0cm, clip=true, width=\textwidth]{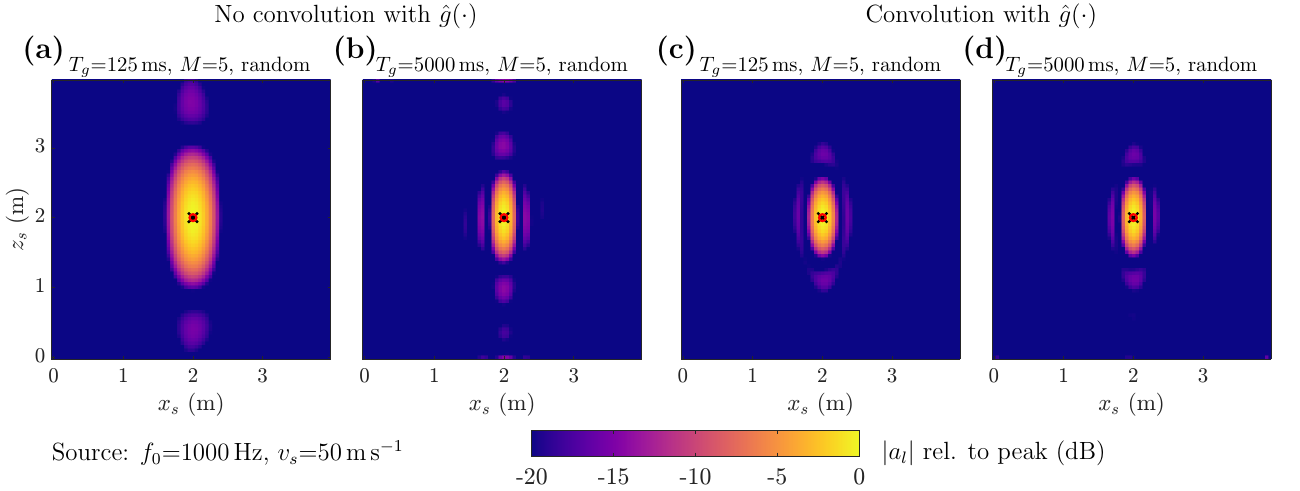}
\caption{Effect of ignoring the spectral leakage on the source maps. Shown are the amplitudes of the regularized solutions for $\Fm=1000$\,Hz 
and $v_s=50$\,m\,s$^{-1}$ using a short window ($T_g=125$\,ms) in panels (a) and (c) and a long window ($T_g=5000$\,ms) in panels (b) and (d). All cases shown used $M=5$ randomly assigned frequency bins per microphone. Panels (a) and (b) show the results without taking the convolution into account in the transfer function, (c) and (d) show the case when the convolution is taken into account. The black x denotes the true source position, the red circle the position of the detected maximum. The dynamic range is set to 20\,dB and the maps are normalized to the respective peak amplitude.
} 
\label{fig:randomfreqwindow}
\end{center}
\end{figure}

The necessity of including the window function is demonstrated in Fig.~\ref{fig:randomfreqwindow}. Figs.~\ref{fig:randomfreqwindow}(c) and (d) show the localization performance of the transfer function including the convolution with the window function, Figs. ~\ref{fig:randomfreqwindow}(a) and (b) without, i.e. the convolution with the window function is replaced by a simple point evaluation and thereby assuming an ``ideal'' Fourier transform  $\hat{g}(\cdot) = \delta(\cdot)$. The DFT is still applied to the ``measured'' time signal. For short time windows (Figs.~\ref{fig:randomfreqwindow}(a) vs. (c)), there is a clear improvement of localization performance, for longer time windows the difference in performance gets smaller because longer time windows are more concentrated around $\omega = 0$ in the frequency domain. 

\subsection{Effect of noise}
Up to now, all examples were in a noise-free setting except for the tiny amount of noise added for better convergence of the L-curve. To investigate the sensitivity of the presented approach with respect to noise, bandpass-filtered Gaussian white noise was used. To test a slightly more realistic setting, the noise was not directly added to each measurement signal, as it would be in the case of typical uncorrelated measurement noise related to the recording equipment. Rather, a stationary point source located at position $(20,10,1)^\top$ was assumed, which creates background noise that was added to the signal at each microphone. This way the noise at the different microphones is correlated, mimicking the case of some noise source nearby. As the source to be detected is moving, the signal-to-noise ratio is defined using the peak amplitude on the first channel of the array (close to the array center) compared to the energy of the stationary noise.

Only the case with $M=5$, $f_0=1000$\,Hz, and $v_s=50$\,m\,s$^{-1}$ was considered in this section. 10 different random frequency sets were generated and 10 different SNRs were considered ranging from 0 to 80\,dB. Note that only a single realization of the noise was used. Together with 7 different values for $T_g$ this resulted in 700 test cases. 

\begin{figure}[!ht]
\begin{center}
\includegraphics[trim=0cm 0cm 0cm 0cm, clip=true, width=0.99\textwidth]{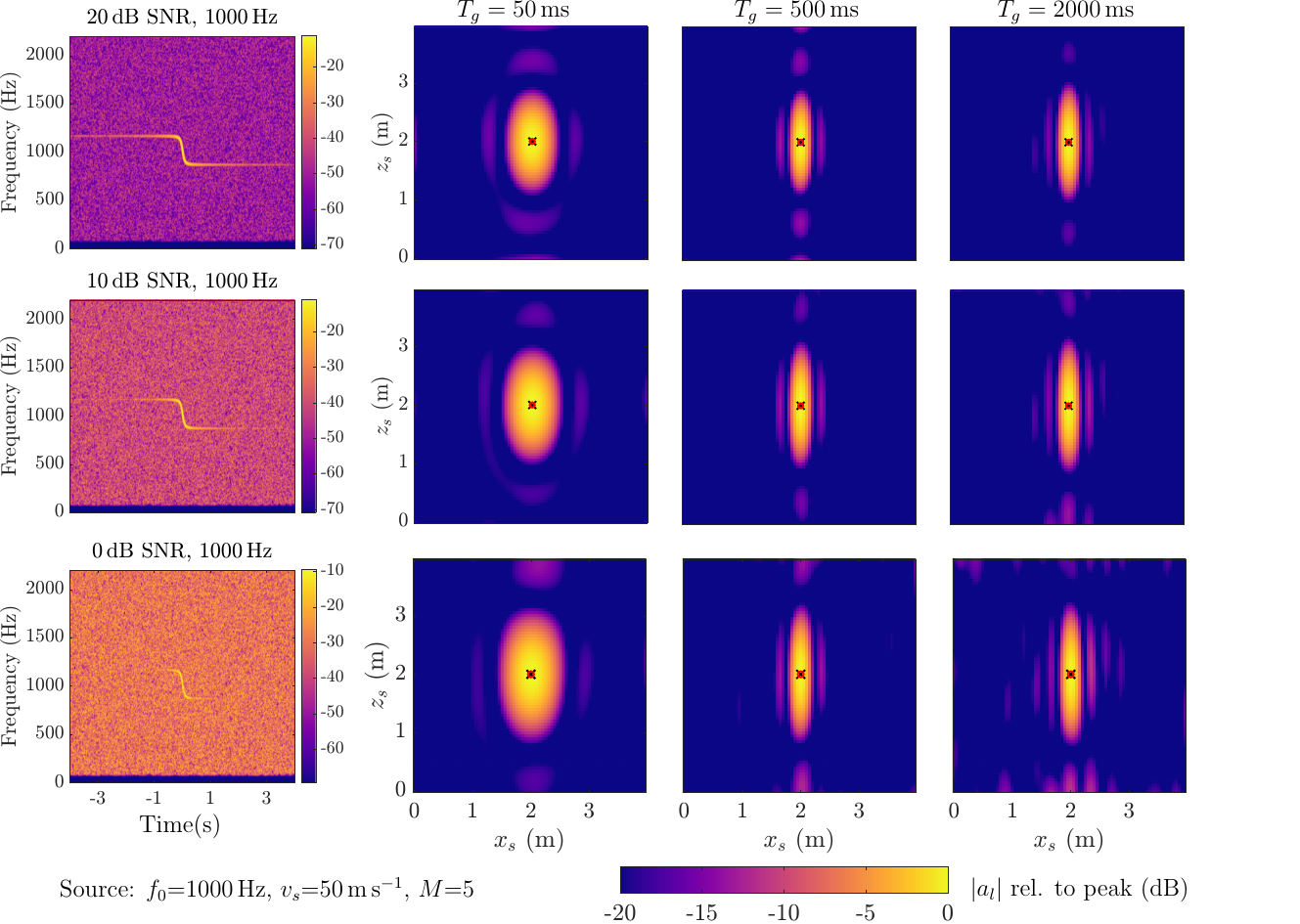}
\caption{Effect of different SNRs on signal and source maps. The left column shows the time-frequency representations of noisy simulated pass-by signals at different noise levels (rows). 
From the second to the fourth column, the analysis window length is increased from 50\,ms (second column) to 2000\,ms (right column).  The black x denotes the true source position, the red circle the position of the detected maximum. The dynamic range is set to 20\,dB and the maps are normalized to the respective peak amplitude.  } 
\label{fig:snrmaps}
\end{center}
\end{figure}

\begin{figure}[!ht]
\begin{center}
\includegraphics[trim=0cm 0cm 0cm 0cm, clip=true, width=0.6\textwidth]{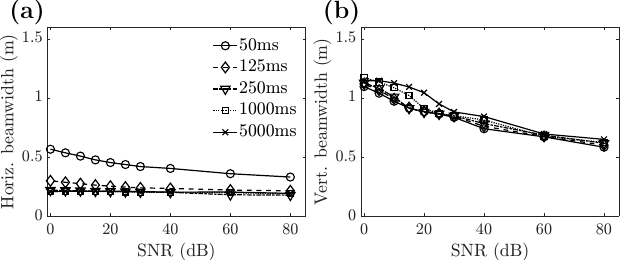}
\caption{Effect of SNR on the beamwidth. The horizontal (a) and the vertical (b) beamwidth are shown as a function of the SNR and the analysis window length $T_g$ (symbols and line type).} 
\label{fig:snreffect}
\end{center}
\end{figure}

The first column in Fig.~\ref{fig:snrmaps} shows the time-frequency representation of the ``recorded'' stimuli with different SNRs (different rows). Plots were generated using the functions \texttt{dgt} and \texttt{plotdgt} of the large time-frequency analysis toolbox (LTFAT, \cite{ltfatnote030}).  It can be seen that the tails in particular are affected by the noise when the source is far away from the microphones. In the remaining columns, the effect of the noise on the source maps is shown for different levels of the SNR and different window lengths. The vertical blurring is more strongly affected by the SNR than the horizontal blurring. Fig.~\ref{fig:snreffect} shows this effect in more detail based on the 3-dB beamwidth as used before. Interestingly, the horizontal beamwidth is only affected by noise for shorter windows. For windows with length $T \ge 250$\,ms, different SNRs do not have an effect on the horizontal beamwidth. In contrast, the vertical beamwidth gets smaller with increasing SNR. For $T_g\geq1000$\,ms, increasing the noise level leads to a deterioration already at lower noise levels compared to shorter windows.

For different random samples, the general trends are essentially the same and thus only the results for a single random selection are shown to avoid cluttering in Fig.~\ref{fig:snreffect}.  The variation of the vertical beamwidth between different random samples is up to 10\,cm, for low SNRs but for SNRs of 20\,dB and higher the variation lies below 5\,cm. The range of the variation of the horizontal beamwidth is at most 2.5\,cm, but mostly it is below 1\,cm. 
The leveling-off of the vertical beamwidth for longer windows, in particular for $T_g=5000$\,ms at an SNR around 0\,dB, is to be treated with caution as the degree of the effect depends on the actual frequencies selected. At high noise levels spurious sidelobes start to appear which increasingly affect the central lobe, as can be seen, for example, in the lower right subfigure in Fig.~\ref{fig:snrmaps}).

The reason for the opposite trend in the horizontal direction is not entirely clear. While the increase for the vertical beamwidth for higher noise levels is to be expected, the robustness of the horizontal beamwidth except for very short windows is somewhat unexpected. As the tendencies are similar for fully uncorrelated noise, the correlation across receivers of the observation noise is not the reason for this effect.
From a detailed analysis of the noisy spectra it is clear that spectra generated by longer or shorter windows are affected differently. For very short windows, the spectral levels decay already relatively close to $f_0$ (c.f. Fig.~\ref{fig:motioneffect}(b)). Thus, the added noise starts to corrupt more peripheral frequency bins at relatively low noise levels while bins close to $f_0$ are much less affected. In contrast, for windows longer than the transition zone, increasing the noise will affect the spectrum much more uniformly. In addition, if the window length is increased further, the noisiness of the spectrum will increase even for a constant SNR. This becomes clear when looking at the time-frequency representation as the signal is highly localized. This explains why for $T_g\geq1000$\,ms increased noise affects the vertical beamwidth much stronger than for shorter windows. The reason for the horizontal beamwidth not being affected is still unclear. However, the appearance of relevant spurious sidelobes is in general problematic. Thus, the better resolution along the direction of motion is of limited practical relevance in high noise cases.

Concerning the accuracy of the localization, the peak coordinate in the source map was in the majority of cases located at the true position. 20 cases were displaced by one grid point (5\,cm) and only two cases had a larger deviation (10 and 15\,cm), but only at very high noise levels. All deviations appeared along the $z$-dimension.

\subsection{Effect of a ground reflection}
\label{Sec:Ground}
Using a numerical method like the 2.5D BEM has the advantage that it allows the inclusion of scattering objects into the source localization, which may otherwise adversely affect the performance of standard beamforming approaches. Although the full BEM framework is not used here, a straightforward example comprising a reflecting half-plane is shown to illustrate the general principle. The half-plane is located 1\,m below the lower edge of the source grid. Thus, the assumed source lies 3\,m above the plane resulting in considerable phase effects due to the combination of the true source and the mirror source caused by the ground plane. The reference data for the moving point source is simply the sum of the moving point source and the mirror source. Similarly, the 2D Green's function for the fully reflecting half-plane is a sum of 2 Hankel functions with different 2D-radii $r_2$. 

\begin{figure}[!ht]
\begin{center}
\includegraphics[trim=0cm 0cm 0cm 0cm, clip=true, width=\textwidth]{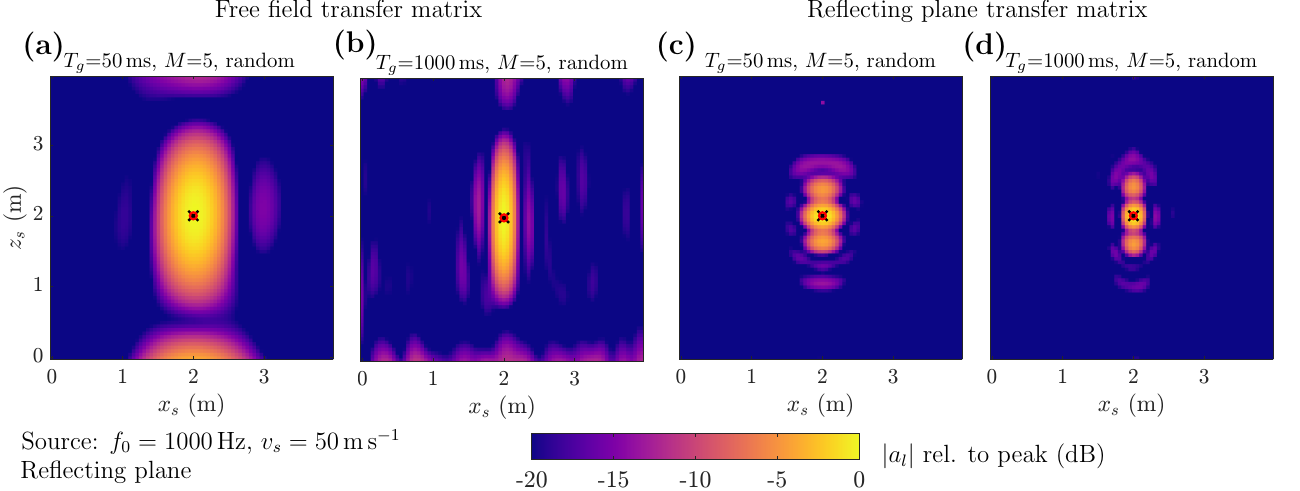}
\caption{Source maps using a different set of $M=5$ random frequencies for each microphone for a moving source above a reflecting half-plane. Shown are the amplitudes of the regularized solutions for $\Fm=1000$\,Hz and $v_s=50$\,$m\,s^{-1}$ using a short window ($T_g=50$\,ms) in panels (a) and (c) and a long window ($T_g=1000$\,ms) in panels (b) and (d). Panels (a) and (b) show the inversion result when the reflecting plane is not considered in the transfer matrix (i.e.the free-field case is used), panels (c) and (d) the case when it is considered. The dynamic range is set to 20\,dB and the maps are normalized to the respective peak amplitude.
} 
\label{fig:randomfreqmirr}
\end{center}
\end{figure}

Fig.~\ref{fig:randomfreqmirr} shows the results for a point source moving above a reflecting half-plane using the same parameters as in Fig.~\ref{fig:randomfreq} with $M = 5$. Figs.~\ref{fig:randomfreqmirr}(a) and (b) show the results when the reflection is only considered in the reference signal at the microphones but not in the transfer matrix, in panels (c) and (d) the ground reflection is also considered when calculating the transfer matrix. If the reflection present in the ``measured'' signal is ignored when generating the transfer matrix, the result is much more blurred (Fig.~\ref{fig:randomfreqmirr}(a) and (b)). 

It is important to note that the L-curve did not work well in this mismatched condition. The corner detected without any limitation of the regularization lead to solutions which had little regularization and the resulting maps comprised mostly heavily scattered blobs or very strong interference patterns (not shown). Only after increasing the degree of regularization the resulting maps were related to the true source distribution. However, due to the tendency of the $L_2$ regularization to distribute the source strengths over a larger number of coefficients (c.f. \ref{App:SingleCase}), this comes at the cost of the heavy blurring shown in Figs.~\ref{fig:randomfreqmirr}(a) and (b).

When considering the reflection in the transfer matrix (panels (c) and (d)), the beamwidth of the main lobe is much smaller, but sidelobes appear in the vertical direction. A closer inspection shows that when including the main sidelobes the overall spread of the source map in panels (c) and (d) is only slightly larger compared to the pure free-field case and that the sidelobes appear in a region similar to the source map identified in the free-field case. 
Along the direction of motion $x$ only minor differences if any can be seen in the spread when comparing the case with (Fig.~\ref{fig:randomfreqmirr}(c) and (d)) and without a reflecting half-plane (Fig.~\ref{fig:randomfreq}(b) and (d)).

\begin{figure}[!ht]
\begin{center}
\includegraphics[trim=0cm 0cm 0cm 0cm, clip=true, width=\textwidth]{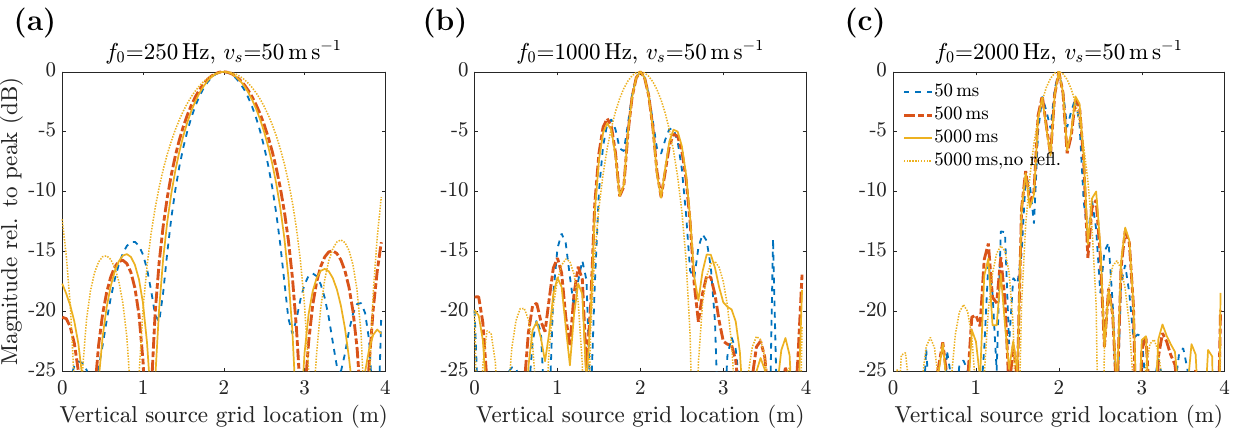}
\caption{Comparison of amplitude profiles of central vertical sections of source maps for the half-plane and the free-field case. The source frequency varies from left to right: 250\,Hz (a), 1000\,Hz (b), and 2000\,Hz (c). Colors and line types denote different window lengths. Thick lines show the half-plane case, the thin dotted line the free-field case. Data were normalized to a peak of 0\,dB.
}
\label{fig:sidelobes}
\end{center}
\end{figure}

Fig.~\ref{fig:sidelobes} illustrates the differences between the half-plane and the free-field case in more detail.
Shown are vertical sections in the center of the respective source maps ($x = 2$\,m) for the half-plane case for different window lengths (dashed, solid, and dash-dotted lines for $T_g = 50, 500,$ and $5000$\,ms). For comparison, the source maps for the pure free-field case with no mismatch are shown for a window length of $T_g = 5000$\,ms (thin dotted line). The mismatched condition is not shown. When defining 3\,dB for the cutoff of the central peak, the beamwidth in $z$ of the central peak is typically lower then in the free-field case. For higher frequencies (Fig.~\ref{fig:sidelobes}(b) and (c)) interference patterns appear along $z$. Comparing the width of the main lobe in the free-field and the combination of main lobe and first sidelobes in the half-plane case shows a similar range of the blurring in both cases.

\section{Discussion}
\label{sec:disc}

The main goal of this study was to investigate the performance of the proposed algorithm using simulated data of single-frequency sources with different fundamental frequencies, speeds, and distances. These artificial data were generated using a time domain formulation which also allowed a direct assessment of the correctness of the frequency domain approach used for inversion. Overall, a good localization performance could be achieved with a placement error of at maximum 1 source grid point in almost all cases.
Thus, the performance lies in a similar range as published results for other methods with at least partially comparable test cases \cite{Cousson2019,Meng2019,Zhang2023}, where differences between true and estimated source position below 0.1\,m were achieved. The direct comparison needs to be interpreted with care as there are differences in the test cases. For example, in \cite{Cousson2019} and \cite{Meng2019} the source speeds considered were only up to about 28\,m\,s$^{-1}$ whereas in \cite{Zhang2023} a speed of 75\,m\,s$^{-1}$ was used. 

 The algorithm allows for the use of long windows, and thus a good spectral resolution even for high source speeds. Moreover, the algorithm even takes advantage of the motion allowing a better resolution along the direction of motion.
A detailed analysis of the results lead to two key parameters affecting the performance: a) The inclusion of the effect of a windowed DFT into the transfer matrix via a convolution with the Fourier transform of the analysis window allows for a much better resolution of the point source, in particular for shorter windows. b) As the motion of the source leads to a spectral range at the receiver points the choice of observation frequency bins used in the comparison between measurement and numerical model has a profound effect on the algorithms performance. 

It was shown that using only one single observation frequency equal to the source frequency for the inversion (as would be the case for stationary sources) lead to a large degree of blurring of the source map along the direction of motion, in particular for higher speeds and longer analysis windows which is caused by a combined effect of increased ambiguities in the transfer functions and the $L_2$ regularization. Using a regular frequency grid to cover the spectral range caused by the motion of the source lead to a better resolution along the direction of motion, but sometimes also induced periodicity in the results caused by a similar effect as the blurring in the single bin case. The periodicity depended on the source speed, the window length, and the chosen frequency spacing and was overcome by using a different set of random observation frequencies for each microphone. 
The horizontal resolution, i.e. the resolution along the direction of motion, could be greatly improved compared to the source frequency setting without inducing  periodicity in the solution. Also, longer analysis windows lead to less horizontal blurring, the effect being dependent on the source speed and frequency. 
However, the use of only one single random observation frequency per microphone lead to a comparatively poor vertical resolution and a high variability across different random choices. Increasing the number of observations by choosing more observation frequencies per microphone yielded a more reliable inversion of the transfer matrix and a decreased vertical blurring which then was similar across a range of source speeds including the almost stationary case. 

The use of different noise levels illustrated that the horizontal blurring was largely unaffected by the signal-to-noise-ratio. In the vertical direction, however, the noise level strongly affected the vertical resolution. 

Results for a reflecting ground showed the versatility of the principle approach and illustrated the importance of including surrounding structures that may affect the propagation. Importantly, the method can be expanded to more complex scenarios. A coupling with numerical methods like the BEM allows for the inclusion of scattering objects as long as the assumption of a constant cross-section for the scatterer along the longitudinal dimension is fulfilled (e.g. a noise barrier).

The main drawback of the method is the high computational effort. In particular, the numerical integration used for this proof-of-concept is far from optimal. Future work beyond the current state will focus on better quadrature approaches based, e.g. on methods for highly oscillatory integrands such as those used in \cite{Kasess2016}.
Also the choice of the measurement frequency range was not optimized for different window lengths and frequencies which will also be considered in the future. Currently, the approach is restricted to single-frequency sources. Multiple such sources with different frequencies can be handled separately, as long as the frequency ranges are not overlapping. For overlapping signals, a joint inversion will be necessary which is ongoing work. Future work will also focus on more general broadband signals. For this purpose, the computational effort required for a more general source signal becomes increasingly important. 
To investigate the effect of model parameters on the resolution an off-the-shelf Tikhonov regularization was used for the inversion employing an L-curve approach to determine the optimal regularization parameter which was stabilized by adding a tiny amount of noise to the observations as suggested in the literature. In the future it is also planned  to incorporate deconvolution methods or the direct use of methods enforcing some degree of sparsity. Current results show that the resolution can be controlled up to a certain degree, which will most likely also benefit sparse approaches. 

\section{Conclusions}
\label{sec:concl}
In this work, an inverse source localization algorithm for single-frequency uniformly moving sources was presented that acts entirely in the frequency domain and utilizes a 2.5D approach. Despite being defined fully in the frequency domain the approach does allow to consider the Doppler exactly without the need to resort to short quasi-stationary segments. Importantly, the effects caused by applying a discrete Fourier transform to the microphone recordings were also included when calculating the transfer functions from potential source points to the microphone positions. 

It was shown that the newly proposed algortithm works particularly well along the direction of motion where the resolution can be significantly improved compared to the elevation of the source. It was also shown that there are two aspects necessary for the good performance of the algorithm: a) The introduction of spectral leakage into the forward 2.5D model, and b) utilizing the spectral broadening, caused by the motion of the source, via randomly distributing different observation frequency bins to different microphones in the recording array.

\section*{Acknowledgements}
This work was supported by the Austrian Science Fund (FWF) via the DACH project LION (Localization and Identification of moving noise sources, I~4299-N32). The authors would like to thank Timo Schumacher from TU~Berlin for valuable discussions about the beamforming perspective on source localization.

\appendix
\section{}
\renewcommand{\theequation}{A.\arabic{equation}}
\subsection{Single frequency bin}\label{App:SingleCase}
If the same single frequency bin is selected for all microphones the localization performance in $x$ deteriorates for large $T_g$. In the following the transfer function for the limiting case $T_g \rightarrow \infty$ is derived. In this setting the window (e.g. a Hanning or a rectangular window) converges towards a function constant in time which has the delta functional as its Fourier transform. Thus, Eq.~(\ref{Equ:Monoxom}) simplifies to
\begin{align}
\nonumber \phat_{n\ell}[\omega'] &=  \frac{a_\ell}{ 2 \pi |v_s|} \sint   \q{n\ell}{\sqrt{\omega^2/c^2-(\omega-\Om)^2/v_s^2}} \hat{g}(\omega'-\omega) \E^{\I (\omega - \Om)v_s^{-1} (x_{r,n} - x_{s,\ell})} d\omega  \\
&=  \frac{a_\ell}{ 2 \pi |v_s|} \q{n\ell}{\sqrt{\omega'^2/c^2-(\omega'-\Om)^2/v_s^2}} \E^{\I (\omega' - \Om)v_s^{-1} (x_{r,n} - x_{s,\ell})} ,
   \label{Equ:Monoxom_single}           
\end{align}
where for $\omega'$ the index $m$ was dropped for simplicity. Strictly speaking, in the limit of $T_g \rightarrow \infty$ the DFT becomes the discrete time FT which is continuous in frequency. To avoid confusion the brackets are kept although it is a slight abuse of notation.

If $\omega'=\Om=2 \pi f_0$, Eq.~(\ref{Equ:Monoxom_single}) reduces to 
\begin{align}
\phat_{n\ell}[\Om] &=  \frac{a_\ell}{ 2 \pi |v_s|} \q{n\ell}{\sqrt{\Om^2/c^2}} = a_\ell \,  h_{n\ell}[\Om]
   \label{Equ:Monoxom_single0}           
\end{align}
where clearly no dependency on the $x$-coordinates of source nor receiver is present in $h_{n\ell}[\Om]$. However, even for $\omega'\neq \Om$ a unique identification of the source position along $x$ is not possible for $T_g \rightarrow \infty$ if the same single $\omega'$ is used for all receivers.  

To illustrate this behavior, a simple test case with a 1-D source grid $\bx_{s,\ell} = (x_{s,0}+\ell \Delta x_s,y_s,z_s)^\top$ and a 1-D receiver grid $\bx_{r,n} = (x_{r,0}+n \Delta x_r,y_r,z_r)^\top$ is used. Using Eq.~(\ref{Equ:Monoxom_single}) and the definition of the transfer function leads to the following expression for $h_{n\ell}[\omega']$:
\begin{align}
h_{n\ell}[\omega'] &=  \frac{1}{ 2 \pi |v_s|} \q{}{\sqrt{\omega'^2/c^2-(\omega'-\Om)^2/v_s^2}} \E^{\I (\omega' - \Om)v_s^{-1} (x_{r,0} +n \Delta x_r - x_{s,0} - \ell \Delta x_s)}.
\label{Equ:Monoxom_1d}           
\end{align}
The indices on $\hat{q}$ were dropped as $\hat{q}$ depends only on the source and receiver positions in the $y$-$z$--plane and is thus independent of $n$ and $\ell$ for the pure 1D case. This expression is split into a part depending on $n$ and one on $\ell$:
\begin{align}
h_{n\ell}[\omega'] &=  h_{n}[\omega'] \E^{- \I (\omega' - \Om)v_s^{-1} \ell \Delta x_s}.
\label{Equ:Monoxom_1d1}           
\end{align}
It is easy to see that the source grid position only leads to a change in the phase of $h_{n\ell}[\omega']$.

Let $\tilde{p}_n[\omega']$ be the observed value on the $n$-th receiver at angular frequency $\omega'$ and $\mathbf{a} = (a_1, \ldots, a_L)^\top$ a solution vector that minimizes the error:
\begin{align}
 \sum\limits_{n=1}^{N}{\left\lVert   \tilde{p}_n[\omega'] - \sum\limits_{\ell=1}^{L}{ a_\ell h_{n\ell}[\omega']} \right\rVert_2^2 } =  \sum\limits_{n=1}^{N}{\left\lVert   \tilde{p}_n[\omega'] - h_n[\omega'] \sum\limits_{\ell=1}^{L}{a_\ell \E^{- \I (\omega' - \Om)v_s^{-1} \ell \Delta x_s}} \right\rVert_2^2 }.
\end{align}
The expression on the right hand side shows that any vector $\mathbf{a}$ leads to the same error, as long as the sum  of its phase shifted coefficients is kept constant. Thus, there is no unique optimal solution. 

To show why this leads to a distributed solution with equal source weight amplitudes the effect of the $L_2$ regularization needs to be taken into account. To simplify the notation, the phase factor is absorbed into some $\tilde{a}_\ell = a_\ell \E^{- \I (\omega' - \Om)v_s^{-1} \ell \Delta x_s}$. Furthermore, for simplicity and without loss of generality, it is assumed that  $\sum\limits_{\ell=1}^{L}{\tilde{a}_\ell} = 1$, thus  $\tilde{a}_L = 1 -  \sum\limits_{\ell=1}^{L-1}{\tilde{a}_\ell}$. With these assumptions the regularization term can now be treated separately.

Using $\tilde{a}_\ell=\alpha_\ell+\I \beta_\ell$, the $L_2$ norm of $\tilde{\mathbf{a}}$ can be written as:
\begin{align}
\nonumber \left\lVert   \tilde{\mathbf{a}}  \right\rVert_2^2 = \sum\limits_{\ell=1}^{L}{\left|\tilde{a}_\ell \right|^2} &= \sum\limits_{\ell=1}^{L-1}{\left|\tilde{a}_\ell \right|^2} + \left| 1 -  \sum\limits_{\ell=1}^{L-1}{\tilde{a}_\ell} \right|^2 \\
&= \sum\limits_{\ell=1}^{L-1}{\alpha_\ell^2} + \left| 1 -  \sum\limits_{\ell=1}^{L-1}{\alpha_\ell} \right|^2 + \sum\limits_{\ell=1}^{L-1}{\beta_\ell^2} + \left| 0 -  \sum\limits_{\ell=1}^{L-1}{\beta_\ell} \right|^2,
\end{align}
where the condition on the sum of the coefficients $\tilde{a}_\ell$ was used to define $\alpha_L$ and $\beta_L$. To find the optimal weighting, the derivatives with respect to the real and imaginary parts of the coefficients are taken separately and set to $0$:
\begin{align}
\nonumber \frac{\partial \left\lVert   \tilde{\mathbf{a}}  \right\rVert_2^2}{\partial \alpha_\ell} &= 2 \alpha_\ell  - 2 \left(  1 -  \sum\limits_{\ell=1}^{L-1}{\alpha_\ell} \right) = 0 \\
\frac{\partial \left\lVert   \tilde{\mathbf{a}}  \right\rVert_2^2}{\partial \beta_\ell} &= 2 \beta_\ell  - 2 \left( 0 -  \sum\limits_{\ell=1}^{L-1}{\beta_\ell} \right) = 0,
\end{align}
 for all $\ell<L$. This leads to $L-1$ equations of the type $\alpha_\ell = 1 - \sum\limits_{\ell=1}^{L-1}{\alpha_\ell} = \alpha_L$ and $L-1$ equations of the type $\beta_\ell = -\sum\limits_{\ell=1}^{L-1}{\beta_\ell} = \beta_L$. Thus, the optimal solution in the $L_2$ sense is given, if all $\tilde{a}_\ell$ for $\ell = 1,\dots,L$ have the same value. In the given example this implies that $\alpha_l=L^{-1}$ and $\beta_\ell=0$ for all $\ell$. For the original coefficients $a_\ell$ this implies the same magnitude but a position-dependent phase factor.
\begin{figure}[!ht]
\begin{center}
\includegraphics[trim=0cm 0cm 0cm 0cm, clip=true, width=0.9\textwidth]{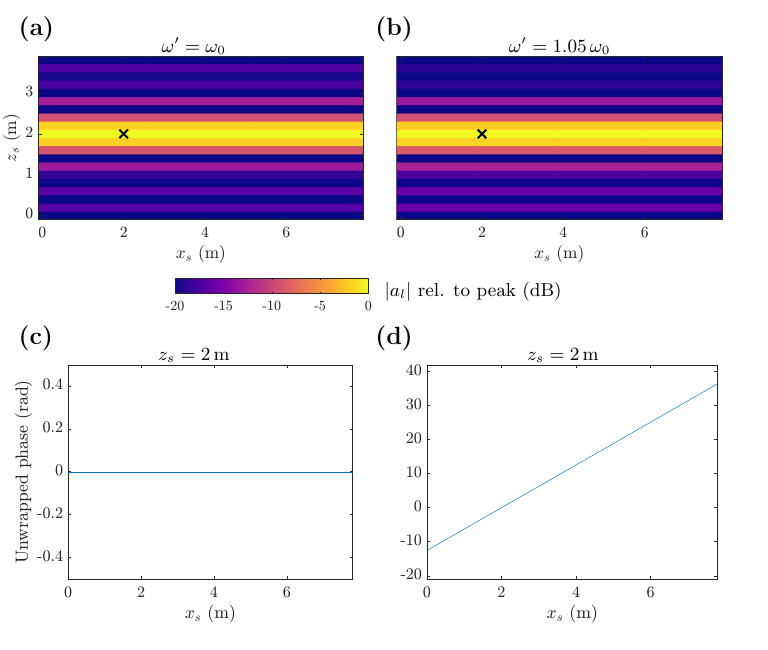}
\caption{Single frequency bin. Shown are the amplitudes of the regularized solution for $\omega'=\omega_0$ (a) and $\omega'= 1.05 \, \omega_0$ (b). Panels (c) and (d) show the unwrapped phase of the complex regularized solution along $x_s$ at the height $z_s=2$\,m for the cases in (a) and (b), respectively. $\omega_0=2000\pi $\,s$^{-1}$ and $v_s=50$\,m\,s$^{-1}$. The black x denotes the true source position.} 
\label{fig:app1}
\end{center}
\end{figure}

To summarize, the smearing along $x$ in the source map is a consequence of the coherent summation of the sources in the transfer functions and the incoherent summation in the regularization term.  In the limit $T_g \rightarrow \infty$ this leads to a complete loss of information about the source along the direction of motion if the same single frequency is considered for all receivers. For finite $T_g$ the leakage effect starts to increase for shorter windows and the results will deviate more strongly from this limiting result.
Importantly, while the example was given for a purely 1D case, this also holds for general 2D receiver grids as the dependence on the source position along $x$ only appears in the exponential term irrespective of the geometrical configuration of the receiver grid. For 2D source grids the treatment of $n$ and $l$ cannot be separated anymore. To illustrate that the general case also leads to similar results, Fig.~\ref{fig:app1} shows the solution for the same scenario used throughout the manuscript using different single frequency bins. However, the transfer functions was calculated for $T_g\rightarrow \infty$. Note that the observations were analyzed for $T_g=10$\,s. The phase (lower plots) shows a clear gradient when $\omega'\neq \Om$ which agrees with the theoretical result of $2 \pi$\,rad\,m$^{-1}$ for the 1D case for $\omega'= 1.05 \Om$. In addition, the phase behaves the same for each line of the source grid along $x$. Thus, the regular grid can be seen as a combination of a number of 1D arrays in $x$. However, due to the joint solution the $L_2$ regularization distributes the energy also along $z$.

\subsection{Regularly spaced frequency bins}\label{App:Regular}
\ref{App:SingleCase} showed how the localization performance in $x$ is destroyed when taking only one single frequency for all receivers. If multiple frequencies in the available frequency range are considered simultaneously, the situation changes.
Similarly to the single-bin case, the regularly spaced case will be treated in the limit $T_g \rightarrow \infty$. For the regular case, the $m$-th frequency of the frequency grid of size $M$ will be defined as $\omega_m' = \omega_s + m \Delta \omega$ which deviates from the definition in the manuscript. 
For example, in Sec.~\ref{Sec:RegBins}, the frequency grid was defined in steps of 50\,Hz starting with 920\,Hz ($M=5$). Thus, a possible choice for the radial frequency grid could be $\omega_s = 2 \pi\, 1020$\,s$^{-1}$ and $m \in \{-2, -1, 0, 1, 2\}$. Combining the regular frequency grid with a regular source and microphone grid, Eq.~(\ref{Equ:Monoxom_1d}) becomes
\begin{align}
h_{n\ell}[\omega_m'] &=  \frac{1}{ 2 \pi |v_s|} \q{}{\sqrt{\omega_m'^2/c^2-(\omega_m'-\Om)^2/v_s^2}} \E^{\I (\omega_s + m \Delta \omega - \Om)v_s^{-1} (x_{r,0} +n \Delta x_r - x_{s,0} - \ell \Delta x_s)},
\label{Equ:Monoxom_1dreg}           
\end{align}
where $\omega_m'$ was only expanded in the exponent to avoid too cluttered an expression. As in Eq.~(\ref{Equ:Monoxom_1d1}) all dependencies on $n$ and $\ell$ can be separated leading to

\begin{align}
h_{n\ell}[\omega_m'] &=  h_{n}[\omega_m'] \E^{- \I (\omega_s - \Om)v_s^{-1} \ell \Delta x_s} \, \E^{- \I  m \Delta \omega v_s^{-1} \ell \Delta x_s}.
\label{Equ:Monoxom_1d1reg}           
\end{align}
The first exponential term has similar consequences as in \ref{App:SingleCase}, i.e. a position dependent phase factor that is independent of  the frequency bin number $m$. The effect of this term has been shown in detail in \ref{App:SingleCase} and is not important for understanding the way periodicity is induced by a regular grid. Thus, for simplicity it is assumed that $\omega_s = \Om$ in the theoretical considerations to follow.  
In contrast to the single bin case, the second phase term affects each frequency bin $\omega_m'$ differently. In the following, it will be illustrated why this leads to a "periodic" solution, which is caused by the the factor
\begin{equation}\label{Equ:Periodfactor}
\E^{- \I \Delta \omega \Delta x_sv_s^{-1} \ell m}
\end{equation}
in the error function. The quotation marks are used, as the solution is, in general,  not truly periodic in the strictest sense.

First, let the frequency and source-grid size be given such that
\begin{align}
\label{Equ:DeltaRel}    
\frac{\Delta \omega \Delta x_s} {v_s} = {2\pi} j,
\end{align}
where $j$ is some integer. For this specific source grid spacing it is clear that $\E^{- \I  \Delta \omega \Delta {x_s} v_s^{-1} m{\ell} } = \E^{- \I  2 \pi m \ell j } = 1$ for any $\ell$ and $m$. Thus, similar to the single bin case, source points, or to be more precise, possible source amplitudes at these points, lying at this particular spacing cannot be uniquely distinguished by the least squares method because possible solutions  only differ by a phase factor when considering the norm of the error:
\begin{align}
  \sum\limits_{n=1}^{N}{
    \sum\limits_{m=1}^{M}{
      \left\lVert
      \tilde{p}_n[\omega'_m] - h_n[\omega'_m] \sum\limits_{{\ell}=1}^{{L}}{a_{{\ell}} }
      \right\lVert
    }
  }.
 \label{Equ:Err_1d1reg}    
\end{align}
Thus the choice of the different components $a_\ell$ is arbitrary as long as the sum over all $\ell$ stays the same. In combination with the condition that $||{\bf a}||$ needs to be small, a similar solution as depicted in Fig~\ref{fig:app1} will be found.

\begin{figure}[!ht]
\begin{center}
\includegraphics[trim=0cm 0cm 0cm 0cm, clip=true, width=0.9\textwidth]{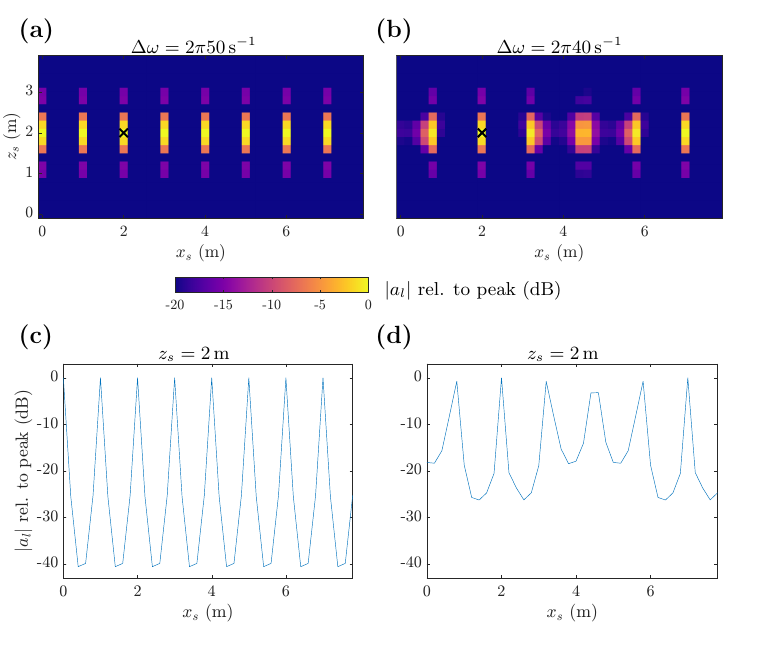}
\caption{Regular frequency bins. Shown are the amplitudes of the regularized solution for $\Delta \omega = 2 \pi 50$\,s$^{-1}$ (a) and $\Delta \omega = 2 \pi 40$\,s$^{-1}$ (b). Panels (c) and (d) show the amplitudes along $x_s$ at the height $z_s=2$\,m for the cases in (a) and (b), respectively. $\omega_0=2\pi 1000$\,s$^{-1}$ and $v_s=50$\,m\,s$^{-1}$. The black x denotes the true source position.} 
\label{fig:app2}
\end{center}
\end{figure}

If $\frac{\Delta \omega \Delta x_s}{v_s} = \frac{2 \pi}j$, where $j$ is some positive integer number, the index set $\mathcal{L} = \{1,\dots, L\}$ can be split into subsets $\mathcal{L}_\alpha = \{\ell: \ell = 1,\dots,L; (\ell \mod  j) = \alpha \}$. In this case, the error functions is given by
\begin{align}\label{Equ:Periodic}
  \sum\limits_{n=1}^{N}{
    \sum\limits_{m=1}^{M}{
      \left\lVert
      \tilde{p}_n[\omega'_m] - h_n[\omega'_m]
      \sum\limits_{\alpha=0}^{\lfloor L/j \rfloor}\E^{-2\pi\I m\alpha/j}
      \sum\limits_{\ell\in \mathcal{L_\alpha}}{a_{\ell}}
      \right\lVert
    }
  },
\end{align}
where $\lfloor \cdot \rfloor$ is denotes the floor of the argument. With the same reasoning as above, the choice of $a_\ell$ in each cluster $\mathcal{L}_\alpha$ is arbitrary. This introduces a periodicity in the error function and thus in the solution of the regularized least square problem as illustrated in Fig.~\ref{fig:app2}(a) which again shows the general case. For this figure $\Delta \omega = 2 \pi 50$\,s$^{-1}$, $\Delta x_s = 0.2$\,m, and $v_s = 50$\, m\,s$^{-1}$, thus $\frac{\Delta \omega \Delta x_s}{v_s} = \frac{2\pi}{5}$. The source was placed at $\bx_0 = [2,0,2]^\top$ with amplitude 1. The $x$ coordinate of the true source, or rather its index in the source grid, lies in the subgroup $\mathcal{L}_0$ which contains all indices $\ell$ with $(\ell \mod 5) = 0$. As only the sum over the indices contained in a subgroup is relevant in Eq.~(\ref{Equ:Periodic}), a strong $L_2$ regularization will weight all $a_\ell$ inside this subgroup equally resulting in the periodic solution with period 5 displayed in Fig.~\ref{fig:app2}(a). Note that a periodic solution will only appear for $L$ being large compared to $j$. 

Even if ${\Delta \omega \Delta x_s}{v_s}^{-1}$ is not of the form $\frac{2\pi}{j}$, a ``periodicity'' can be observed as illustrated in Fig.~\ref{fig:app2}.  For this figure $\Delta \omega = 2\pi 40$\,s$^{-1}$, the rest of the setting remains the same. Note that for this choice of frequencies $M=6$ and $\omega_s$ can be set to $\Om$ with $m\in\{-2 \cdots 3\}$. Now $\frac{\Delta \omega \Delta x_s}{v_s}$ is not of the form $\frac{2\pi}{j}$, however, the solution still looks periodic. As the true source is still on a spatial grid point, the ``true'' error function reads as
\begin{align}
  \sum\limits_{n=1}^{N}{
    \sum\limits_{m=1}^{M}{
      \left\lVert
      \tilde{p}_n[\omega'_m] - h_n[\omega'_m] \E^{2\pi \I \frac{4}{25}11 m  }
      \right\lVert
    }
  } = 0.
\end{align} 
as the source strength for $\tilde{p}$ was set to 1. There is still a true periodicity in the solution because $19 = (4 \cdot 11 \mod 25) = (4 \cdot 36\mod 25)$ and thus, the exponential factor from Eq.~(\ref{Equ:Periodfactor}) is the same for $\ell = 11$ and $\ell = 36$ ($x_s=2$\,m and $x_s=7$\,m, respectively). However, the error function will have local minima for $\ell$-values with $(4 \ell \mod 25)$ close to 19. As a consequence, the $L_2$ regularized least square method finds possible sources around $\ell = 5, 17, 23, 24,$ and $30$ and distributes the energy around those points. If $\ell$ were allowed to be non-integer, the exact positions were $(\ell \cdot 11 \mod 25) = 19$ would be  $\ell = 4.75, 11, 17.25, 23.5, 29.75$ and $36$ yielding positions of $x_s = 0.75, 2, 3.25, 4.5, 5.75,$ and $7$\,m which matches a period of $2 \pi \Delta \omega^{-1} v_s = 50/40 = 1.25$\,m obtained by Eq.~\ref{Equ:DeltaRel} for $j=1$. 

Thus, very loosely speaking, the true periodicity (up to a phase factor) given by the choice of $\Delta \omega$ is either matched if source points are aligned exactly with this induced period or approximated by neighboring points in case of a spatial mismatch.

 \bibliographystyle{elsarticle-num}

\end{document}